\documentclass[journal]{IEEEtran}

\usepackage{enumerate}
\usepackage{times}
\usepackage{hyperref}
\usepackage{breakurl}
\usepackage[final]{graphicx}
\usepackage[reqno]{amsmath}
\usepackage{amsfonts}
\usepackage{multicol}
\usepackage{times,amsmath,epsfig}
\usepackage{latexsym,amssymb}
\usepackage{cite}
\usepackage{subfigure}
\usepackage{amsmath}
\usepackage{cases}
\usepackage{amssymb}
\usepackage{array}
\usepackage[noend]{algorithmic}
\usepackage{algorithm}
\usepackage{color}
\usepackage{psfrag}
\usepackage{epsfig}

\newtheorem{lemma}{Lemma}
\newtheorem{pro}{Proposition}

\newtheorem{theorem}{Theorem}

\newcommand{\mb}{\mathbf}
\newcommand{\mc}{\mathcal}

\newcommand{\E}{\mathbb{E}}

\newcommand{\1}{\mathbf{1}}
\newcommand{\lta}{\lim_{T\to\infty} \frac{1}{T}\sum_{t=0}^{T-1}}

\newcommand{\p}{\mathbb{P}}
\newcommand{\define}{{\triangleq}}

\newcommand{\up}{\textrm{up}}
\newcommand{\ie}{\emph{i.e.}}

\allowdisplaybreaks

\begin{document}

\title{Real-Time Welfare-Maximizing Regulation Allocation in Dynamic Aggregator-EVs System}
\author{Sun Sun, \IEEEmembership{Student Member,~IEEE}, Min Dong, \IEEEmembership{Senior Member,~IEEE}, and Ben Liang, \IEEEmembership{Senior Member,~IEEE}
\thanks{This work was supported in part by the Natural Sciences and Engineering Research Council of Canada.}
\thanks{Sun Sun and Ben Liang are with the Department of  Electrical and Computer
Engineering, University of Toronto, Toronto, Canada (email: \{ssun,
liang\}@comm.utoronto.ca).}
\thanks{Min Dong is with the Department of Electrical Computer and Software Engineering, University of Ontario Institute of Technology, Toronto, Canada (email:  min.dong@uoit.ca).}}

\maketitle

\begin{abstract}
The concept of vehicle-to-grid (V2G) has gained recent interest as more and more electric vehicles (EVs) are put to use. In this paper, we consider a dynamic aggregator-EVs system, where an aggregator centrally coordinates a large number of  dynamic EVs to provide regulation service.  We propose a Welfare-Maximizing Regulation Allocation (WMRA) algorithm for the aggregator to fairly allocate the regulation amount among the EVs.  Compared with previous works, WMRA accommodates a wide spectrum of vital system characteristics, including dynamics of EV,  limited EV battery size, 
EV battery degradation cost, and  the cost of using external energy sources for the aggregator.
The algorithm operates in real time and does not require any prior knowledge of the statistical information of the system. Theoretically, we demonstrate that WMRA is away from the optimum by $O(1/V)$, where $V$ is a controlling parameter depending on EVs' battery size. In addition, 
 our simulation results indicate that WMRA can substantially outperform a suboptimal greedy algorithm.
\end{abstract}


\section{Introduction}
Electrification of personal transportation is expected to become prevalent in the near future.  
For example, from one report of the U.S. department of energy\cite{usengy}, the government sets an ambitious goal  to put one million EVs on the road by 2015.
 Besides serving the purpose of transportation,  EVs can also be used as  distributed electricity generation/storage devices when plugged-in \cite{gui09}.
Hence, the concept of
vehicle-to-grid (V2G), referring to the integration of EVs to the power grid,
has received increasing attention \cite{kem05,gui09}.

Frequency regulation service is to balance  power generation and load demand in a short time scale, so as  to maintain the frequency of a power grid at its nominal value.
Traditionally, regulation service is provided by fast responsive  generators, 
which vary their output to alleviate power deficits or surpluses, 
 and is the most expensive ancillary service \cite{kir05}. 
Experiments show that EV's power electronics and battery can well respond to the frequent regulation signals. Thus it is possible to exploit a plugged-in EV as a promising alternative to provide regulation service through charging/discharging, which potentially could reduce the cost of regulation service significantly \cite{kem08}.
However, 
since  the regulation service is generally requested on the order of megawatts  while  the power capacity of an EV is typically $5$-$20$ kW, it is often necessary for an aggregator to coordinate a large number of EVs  to provide regulation service \cite{bes11}. In addition, frequent charging/discharging has a detrimental effect on EV's battery life. Thus, it is important to design proper algorithm for regulation allocation in the aggregator-EVs system, especially in a real-time fashion.

There is a growing body of recent works on   V2G regulation service.
Specific to the aggregator-EVs system, which focuses on the interaction between the aggregator and  the EVs,
centralized  regulation allocation   is  studied in \cite{gar12,han11b,sor12,han10,shi12}, where the objective is to maximize the profit of the aggregator or the EVs. In \cite{gar12}, a set of schemes based on different criteria of fairness among the EVs are provided.  In \cite{han11b}, the regulation allocation problem is formulated as quadratic programming.  
In \cite{sor12},  considering both  regulation service and spinning reserves,  the underlying  problem is formulated  as linear programming. In  \cite{han10}, the charging behavior of EVs is also considered, and  the underlying  problem is  then reduced to the control of  charging sequence and  charging rate of each EV, which is solved by  dynamic programming.
In \cite{shi12},
a real-time regulation control algorithm is proposed  by formulating the problem as a Markov decision process, with the action space consisting of charging, discharging, and regulation.  Finally,
a distributed regulation allocation system is proposed in \cite{wu12} using game theory, and a smart pricing policy is developed to incentivize  EVs.

In addressing the regulation allocation problem, however,  these earlier works have omitted to consider some essential characteristics of the aggregator-EVs system.  For example, deterministic model is used in \cite{gar12} and \cite{han10}, which ignore the uncertainty of the system, e.g., the uncertainty of the electricity prices. The dynamics of the regulation signals is not incorporated in \cite{wu12}, nor  the energy restriction of EV battery is considered.  The self-charging/discharging activities in support of  EV's own need are omitted in \cite{gar12} and \cite{wu12}.  
The potential cost of using external energy sources for the aggregator to  accomplish regulation service is ignored in \cite{gar12,shi12,han11b,sor12,han10}, and the cost of EV battery degradation due to frequent charing/discharging in  regulation service is not considered in \cite{wu12,shi12,han11b,han10}.

In this work,
we consider all of the above factors in a more complete aggregator-EVs system model, and develop a real-time algorithm for the aggregator to fairly allocate the regulation amount among the EVs.  
Specifically,   considering an aggregator-EVs system providing long-term regulation service to a power grid,  we aim to maximize the  long-term social welfare of the aggregator-EVs system, with the constraints on each EV's regulation amount and degradation cost. To solve such a stochastic optimization problem, we adopt Lyapunov optimization technique, which  is also used in 
\cite{neely10, chen12,   hua13}
 for  demand side management  in smart grid. 
We demonstrate how a solution to this  maximization can be formulated under a general Lyapunov optimization framework \cite{bkneely}, and  propose a real-time allocation strategy specific to the aggregator-EVs system.  The proposed Welfare-Maximizing Regulation Allocation (WMRA) algorithm does not rely on
 any statistical information of the system, and  is shown to  be asymptotically  close to the optimum as  EV's battery capacity increases.  Finally, WMRA is compared to a greedy algorithm through simulation and is shown to offer substantial performance gains.

In our preliminary version of this work \cite{sun13}, the EVs are ideally assumed to be static, \ie, they are  in the aggregator-EVs system throughout the operational time. 
 In this paper,
  to  more realistically capture the dynamics of the aggregator-EVs system, we   generalize the  system model in \cite{sun13} to accommodate dynamic  EVs, which
  is considered in none of the previous works\cite{gar12,han11b,sor12,han10,shi12,wu12}. 
 This generalization
 is challenging for the centralized control of regulation allocation, since the returning EV may have a different energy state compared with the last leaving energy state,  and this energy difference will impose much more difficulties on
 the aggregator for handling  EV's battery size constraint. To tackle this difficulty, we design a novel virtual queue to track the energy state of each EV. Through a careful design of the dynamics of the virtual queue, we can ensure  that the battery size constraint of the EV is always satisfied.

The remainder of this paper is organized as follows. We describe the system model and formulate the regulation allocation problem in Section \ref{sec-sys}. In Section \ref{sec-wmra},  we propose WMRA, and   in Section \ref{sec-perform} we analyze its performance. Simulations  are exhibited in Section \ref{sec-sim}, and we conclude in Section \ref{sec-con}.

\textbf{Notation}: Denote $[a]^{+}$ as $\max\{a,0\}$, $[a,b]^+$ as $\max\{a,b\}$,
and $[a,b]^-$ as $\min\{a,b\}$.  The main symbols used in this paper are summarized in Table \ref{tab:allsym}.

\begin{table}[t]\label{tab:num}
\renewcommand{\arraystretch}{1.5}
\caption{List of Main Symbols}\label{tab:allsym}
\centering
\begin{tabular}{|p{0.7cm}|p{7.2cm}|}
\hline
$G_t$ & regulation signal at time slot $t$ \\
\hline
$\1_{d,t}$ & indicator of regulation down at time slot $t$ \\
\hline
$\1_{u,t}$ & indicator of regulation up at time slot $t$ \\
\hline
$\Delta t$ & interval of regulation signals\\
\hline
$N$ &  number of registered EVs\\
\hline
$t_{ir,k}$ & $k$-th returning time slot of the $i$-th EV\\
\hline
$t_{il,k}$ & $k$-th leaving time slot of the $i$-th EV\\
\hline
$\mathcal{T}_{i,r}$ & set of returning time slots for the $i$-th EV\\
\hline
$\mathcal{T}_{i,l}$ & set of leaving time slots for the $i$-th EV\\
\hline
$\mathcal{T}_{i,p}$ & set of all participating time slots for the $i$-th EV\\
\hline
$\1_{i,t}$ & indicator of  the $i$-th EV's dynamics  at time slot $t$\\
\hline
$x_{id,t}$ & regulation down amount of the $i$-th EV at time slot $t$\\
\hline
$x_{iu,t}$ & regulation up amount of the $i$-th EV  at time slot $t$\\
\hline 
$x_{i,\max}$ & upper bound on  $x_{id,t}$ and $x_{iu,t}$\\
\hline
$x_{i,t}$ & regulation amount of the $i$-th EV  at time slot $t$\\
\hline
$h_{id,t}$ & effective upper bound on $x_{id,t}$\\
\hline
$h_{iu,t}$ & effective upper bound on $x_{iu,t}$\\
\hline
$s_{i,t}$ & energy state of  the $i$-th EV  at the beginning of time slot $t$\\
\hline
$s_{i,\textrm{cap}}$ & battery capacity of  the $i$-th EV\\
\hline
$s_{i,\min}$ & lower bound on $s_{i,t}$\\
\hline
$s_{i,\max}$ & upper bound on $s_{i,t}$\\
\hline
$\Delta_{i,k}$ & difference between the $i$-th EV's $(k+1)$-th returning  energy state and the $k$-th leaving energy state\\
\hline
$C_i(\cdot)$ & degradation cost function  of  the $i$-th EV\\
\hline
$c_{i,\max}$ & upper bound on $C_i(\cdot)$\\
\hline
$c_{i,\textrm{up}}$ & upper bound on long-term degradation cost of  the $i$-th EV\\
\hline
$e_{s,t}$ & unit cost of clearing energy surplus\\
\hline
$e_{d,t}$ & unit cost of clearing energy deficit\\
\hline
$e_{\min}$ & lower bound on $e_{s,t}$ and $e_{d,t}$\\
\hline
$e_{\max}$ & upper bound on $e_{s,t}$ and $e_{d,t}$\\
\hline
$\omega_i$ & normalized weight of the $i$-th EV\\
\hline
\end{tabular}
\vspace{-0.2cm}
\end{table}

\vspace{-0.2cm}

\section{System Model and Problem Formulation}\label{sec-sys}
In this section, we  propose a centralized dynamic aggregator-EVs system and formulate  the regulation allocation problem mathematically.

\vspace{-0.2cm}

\subsection{Aggregator-EVs System and Regulation Service}
Consider a long-term time-slotted system, in which the regulation service is provided over equal time intervals of length $\Delta t$.
At the beginning of each time slot $t\in \mc{T} \define \{0,1,\cdots\}$, the aggregator receives a random regulation signal $G_t$ from a power grid.  If ${G}_t>0$,  the aggregator is required to provide \emph{regulation down} service by absorbing $G_t$ units of energy from the power grid during time slot $t$; if ${G}_t <0$,  the aggregator is required to provide \emph{regulation up} service  by contributing $|G_t|$ units of energy to the power grid during time slot $t$. To represent the type of the regulation service at time slot $t$, we define the indicator random variables $\1_{d,t}\define
\begin{cases}
1 ,& \textrm{ if }G_t >0 \\
0, & \textrm{ otherwise }
\end{cases}$ and
$\1_{u,t}\define
\begin{cases}
1 ,& \textrm{ if }G_t <0 \\
0, & \textrm{ otherwise }
\end{cases}$. 
Note that the product $\1_{d,t}\cdot\1_{u,t}=0$,  since regulation down and up services cannot happen simultaneously.

To provide  regulation service,
the aggregator coordinates $N$ registered EVs and can communicate with each  EV bi-directionally when the EV is  plugged-in.
Each EV can leave  the system for personal reason or for self-charging/discharging purpose and re-join the system later. Assume that each EV provides regulation service only if it is in the system.

For the $i$-th EV,
denote $t_{ir,k}\in \mc{T}$ as its $k$-th returning time slot and $t_{il,k}\in \mc{T}$ as its $k$-th leaving time slot with $t_{ir,k} < t_{il,k}$, $ \forall k\in \{1,2,\cdots\}$. For simplicity of analysis, assume that all EVs are in the system at the initial time  and thus $t_{ir,1} = 0,\forall i$.
Define the set of the returning time slots of the $i$-th EV as
$
\mc{T}_{i,r}\define \{t_{ir,1},t_{ir,2}, \cdots\}
$
 and the set of its leaving time slots as
$
\mc{T}_{i,l}\define \{t_{il,1}, t_{il,2}, \cdots\},
$ with $t_{ir,k}<t_{ir,k+1}$ and $t_{il,k}<t_{il,k+1}$.
Define
$$
\mc{T}_{i,p}\define \cup_{k=1}^{\infty}\{t_{ir,k}, t_{ir,k}+1,\cdots, t_{il,k}-1\}
$$
as the set containing all participating time slots of the $i$-th EV for regulation service. Hence, 
the $i$-th EV  is in the system  for any $t\in \mc{T}_{i,p}$.  Define the
 indicator random variable
$
\1_{i,t}\define
\begin{cases}
1 ,& \textrm{ if }t\in \mc{T}_{i,p}\\
0, & \textrm{ otherwise }
\end{cases}
$ to represent the dynamics of the $i$-th EV  at time slot $t$ (\ie, whether the $i$-th EV is in the system at time slot $t$).
Define  the vector $\1_t\define [\1_{1,t},\cdots, \1_{N,t}]$ to represent  the dynamics of all EVs at time slot $t$.

At the beginning of each time slot,
 the aggregator  allocates  regulation  amount   among all \emph{participating} EVs.   Denote  $x_{id,t}\ge 0$ as the amount of regulation down energy allocated  to the $i$-th EV through charging, and  $x_{iu,t}\ge 0$ as the amount of  regulation up energy contributed by the $i$-th EV through discharging.  Due to the limitation of  charging/discharging circuit in  battery, assume that $x_{id,t}$ and $x_{iu,t}$  are upper  bounded by $x_{i,\max}>0$. Note that if the $i$-th EV is out of the system at time slot $t$, \ie, $\1_{i,t} = 0$, then it cannot provide regulation service and  we have  $x_{id,t}=x_{iu,t}=0$.
 Define the  vectors $\mb{x}_{d,t}\define [x_{1d,t},\cdots,x_{Nd,t}]$ and $\mb{x}_{u,t}\define [x_{1u,t},\cdots,x_{Nu,t}]$ to represent the  regulation amounts of all EVs at time slot $t$.

For the $i$-th EV, assume that it is in the system at time slot $t$ (\ie, $\1_{i,t} = 1$),  and thus can provide regulation service.
Denote $s_{i,t}\in [0, s_{i,\textrm{cap}}]$   as its energy state at the beginning of  time slot $t$, with
  $s_{i,\textrm{cap}}$ being its battery capacity.
Due to  the regulation service,
the energy state of
the $i$-th EV  at the beginning of time slot $t+1$ is given by
 \begin{align}\label{sev}
s_{i,t+1}
= s_{i,t} +\1_{d,t}x_{id,t} - \1_{u,t}x_{iu,t}
=s_{i,t} +b_{i,t},\end{align}
where
\begin{align}\label{bit}
b_{i,t} \define \1_{d,t}x_{id,t} - \1_{u,t}x_{iu,t}
\end{align}
is defined to be the effective charging/discharging amount of the $i$-th EV at time slot $t$.
Charging a battery  near its capacity or discharging it  close to the zero energy state  can significantly reduce battery's lifetime \cite{han11c}.  Therefore, lower and upper bounds on the battery energy state are usually imposed by its manufacturer or user.   Denote the interval $[s_{i,\min}, s_{i,\max}]$ as the preferred energy range of the $i$-th EV with $0\le s_{i,\min}<s_{i,\max}\le s_{i,\textrm{cap}}$. Then,
 the resultant  energy state $s_{i,t+1}$ in \eqref{sev} should lie in  $[s_{i,\min}, s_{i,\max}]$,
which indicates that
the  regulation amounts $x_{id,t}$ and $x_{iu,t}$   must satisfy
$0\le x_{id,t}\le \1_{i,t}h_{id,t}$ and $0\le x_{iu,t}\le \1_{i,t}h_{iu,t}$, respectively,
where $h_{id,t}$ and $h_{iu,t}$ are effective upper bounds on the regulation amounts and are defined as
\[h_{id,t}\define
\left[x_{i,\max}, s_{i,\max}-s_{i,t}\right]^-,
\] and
\[h_{iu,t}\define
\left[x_{i,\max}, s_{i,t}-s_{i,\min}\right]^-,\]
respectively.

From time to time, the $i$-th  EV may need to stop its regulation service and leave the system. When the EV is out of the system (\ie, $\mb{1}_{i,t} = 0$), it cannot offer  regulation service and the aggregator has no information of the EV's energy state. Moreover, the dynamics of the energy state may not follow \eqref{sev} when $\mb{1}_{i,t} = 0$.
When returning,
  the EV may have a different energy state compared with its last leaving energy state. Assume that all returning  energy states  of the $i$-th EV  are confined  in the preferred energy range by the EV's self-control, \ie,  $s_{i,t}\in [s_{i,\min}, s_{i,\max}], \forall t\in \mc{T}_{i,r}$. Define
\begin{align}\label{delta}
\Delta_{i,k} \define s_{i,t_{ir,k+1}}-s_{i,t_{il,k}}, \forall k\in\{1,2,\cdots\}
\end{align}
 as the  difference between the $i$-th EV's  $(k+1)$-th returning energy state and its last leaving energy state.  We assume that 
 \begin{enumerate}
 \item[A1)] $\Delta_{i,k}$ is bounded, \ie, $|\Delta_{i,k}|\le \Delta_{i,\max}$, where the constant $\Delta_{i,\max}\ge 0$. 
 \item[A2)] $\Delta_{i,k}$ has mean zero, \ie,  $\E[\Delta_{i,k}]=0$, $\forall k$.
  \end{enumerate}
  Note that A2 is a mild assumption, based on the random behavior of each EV  when it is out of the system.

For each EV, providing regulation service incurs battery degradation
 due to frequent charging/discharging activities.
Denote $C_i(x)$ as the degradation cost function of the regulation amount of the $i$-th EV, with $0\le C_i(x) \le c_{i,\max}$ and $C_i(0) = 0$.  Since faster charging or discharging, \ie, larger value of $x_{id,t}$ or $x_{iu,t}$, has a more detrimental effect on the battery's lifetime,  we assume   $C_i(x)$ to be convex, continuous, and non-decreasing on the interval $[0, x_{i,\max}]$.
We further assume that each EV imposes an upper bound $c_{i,\textrm{up}}\in [0, c_{i,\max}]$ on the time-averaged battery degradation,
expressed by
\[
\lim_{T\to\infty} \frac{1}{T}\sum_{t=0}^{T-1} \E\left[\1_{d,t}C_i(x_{id,t}) + \1_{u,t}C_i(x_{iu,t}) \right] \le c_{i,\textrm{up}}.
\]

The total regulation amount provided by the EVs may  be insufficient to meet the requested regulation amount due to, for example, a lack of participating EVs, or high battery degradation cost.  For brevity, define
\[x_{i,t}\define \1_{d,t}x_{id,t} +\1_{u,t}x_{iu,t}, \quad 0\le x_{i,t}\le x_{i,\max}\] as the  regulation amount allocated to the $i$-th EV at time slot $t$, which  equals either $x_{id,t}$ or $x_{iu,t}$. Then, the insufficiency of the regulation amount is indicated by $\sum_{i=1}^Nx_{i,t}<|G_t|$, with the   gap  $|G_t|-\sum_{i=1}^Nx_{i,t}$ representing an energy surplus in the case of regulation down or an energy deficit in the case of regulation up. Assume that  energy  surplus or energy deficit must be cleared, or the regulation service fails.  Therefore, from time to time, the aggregator  has to  exploit more expensive external energy sources, such as from the traditional regulation market, so as to fill the energy gap. Denote the unit costs for clearing energy surplus and energy deficit at  time slot $t$ as $e_{s,t}$ and $e_{d,t}$,
respectively, which are both random but are restricted in the interval $[e_{\min}, e_{\max}]$.  Then, the  cost of the aggregator for using the external  energy sources at  time slot $t$  is given by
\begin{align*}
e_t \define  \1_{d,t}e_{s,t}\Big(G_t-\sum_{i=1}^Nx_{id,t}\Big)
+ \1_{u,t}e_{d,t}\Big(|G_t|-\sum_{i=1}^Nx_{iu,t}\Big),
\end{align*}
where we have implicitly assumed that the total regulation amount provided by all EVs cannot exceed the requested amount.

\subsection{Fair Regulation Allocation through Welfare Maximization}
The objective of the aggregator  is to maximize the long-term social welfare of the aggregator-EVs system. Specifically, the aggregator aims to fairly allocate the regulation amount among  EVs and to reduce the cost  for  the expensive external energy sources, with the constraints on each EV's regulation amount and degradation cost.
To this end, we formulate the regulation allocation problem as  the following stochastic optimization problem\footnote{For EVs that only visit the system finite times, since they only affect the system's transient behavior, but not the long-term behavior, we can ignore them and only consider the rest EVs that leave and re-join the system infinite times.}:
\\
{\bf P1: }
\begin{align}
\nonumber
\hspace{-0.2cm}\max_{\mb{x}_{d,t}, \mb{x}_{u,t}} & \sum_{i=1}^N \omega_iU\big(\lim_{T\to\infty} \frac{1}{T}\sum_{t=0}^{T-1} \E[x_{i,t}]\big)
-\lim_{T\to \infty}\frac{1}{T}\sum_{t=0}^{T-1}\E[e_t]\\
\label{p1xdcst}
\textrm{s.t.} \quad & 0 \le x_{id,t} \le \1_{i,t}h_{id,t}, \quad \forall i,t\\
\label{p1xucst}
& 0 \le x_{iu,t} \le \1_{i,t}h_{iu,t}, \quad \forall i,t\\
\label{p1dtcst}
&  \sum_{i=1}^Nx_{id,t} \le \1_{d,t}G_t,\quad \forall t\\
\label{p1utcst}
&  \sum_{i=1}^Nx_{iu,t} \le \1_{u,t}|G_t|, \;\;\forall t\\
\label{p1ccst}
&
\hspace{-1.2cm} \lim_{T\to\infty} \frac{1}{T}\sum_{t=0}^{T-1} \E\left[\1_{d,t}C_i(x_{id,t}) + \1_{u,t}C_i(x_{iu,t}) \right] \le c_{i,\textrm{up}},
\forall i,
\end{align}
where $\omega_i>0$ is  the normalized weight associated with  the $i$-th EV, and
$U(\cdot)$ is a utility function assumed to be  concave, continuous, and non-decreasing, with $U(0)=0$.
Furthermore, to facilitate later analysis, we make a mild assumption that the utility function $U(\cdot)$ satisfies
\begin{align}\label{ucond}
U(x)\le U(0) + \mu x,  \forall x\in\left[0, \max_{1\le i\le N}\{x_{i,\max}\}\right],
\end{align}
where the constant $\mu>0$. One sufficient condition for (\ref{ucond}) to hold is that $U(\cdot)$  has finite positive derivate at zero, such as $U(x) = \log(1+x)$.
The expectations in the above optimization problem  are taken over the randomness of the system  and the possible randomness of the regulation allocation.

In the objective function of {\bf P1}, the first term includes each EV's welfare under the  utility function $U(\cdot)$ and the weight $\omega_i$, and the second term reflects the aggregator's cost for exploiting external energy sources. 
Note that the fairness of the regulation allocation among EVs is ensured by the utility function $U(\cdot)$, and various types of fairness can be achieved by using different utility functions \cite{bknoc}.  For each EV,  in  \eqref{p1xdcst} and \eqref{p1xucst},  hard constraints on  the regulation amounts are set at each time slot, while in \eqref{p1ccst}, a long-term time-averaged constraint  on the regulation amount  is set due to the battery degradation.
The constraints  (\ref{p1dtcst}) and (\ref{p1utcst}) ensure that $x_{id,t} =0$ for regulation up and $x_{iu,t} =0$ for regulation down.

\vspace{-0.2cm}

\section{Welfare-Maximizing Regulation Allocation}\label{sec-wmra}
In this section, we  first apply a sequence of two reformulations to \textbf{P1}, then
propose a real-time  welfare-maximizing regulation allocation (WMRA) algorithm to solve the resultant optimization problem. The performance analysis of
 the proposed WMRA  will be shown in Section \ref{sec-perform}.

\vspace{-0.2cm}
\subsection{Problem Transformation}\label{subsec-protran}
The objective of \textbf{P1} contains  a function of a long-term time  average, which complicates the problem.  Fortunately, in general, such a problem can be transformed to a problem of maximizing the long-term time average of the function \cite{bkneely}.  Specifically, we transform \textbf{P1} as follows.

We first introduce an auxiliary  vector $\mb{z}_t \define [z_{1,t}, \cdots, z_{N,t}]$ with the constraints
\begin{align}
\label{p2aux0}
&0\le z_{i,t}\le x_{i,\max}, \forall i, t, \textrm{ and }\\
\label{p2aux}
& \lim_{T\to\infty} \frac{1}{T}\sum_{t=0}^{T-1} \E[z_{i,t}] = \lim_{T\to\infty} \frac{1}{T}\sum_{t=0}^{T-1} \E[x_{i,t}],\forall i.
\end{align}
From the above constraints, the auxiliary variable $z_{i,t}$  and the regulation allocation amount $x_{i,t}$ lie in the same range and have the same long-term  time average behavior.
We next consider the following problem.
\\
{\bf P2: }
\begin{align*}
&\max_{\mb{x}_{d,t},\mb{x}_{u,t},\mb{z}_{t}}  &
\lim_{T\to\infty} \frac{1}{T}\sum_{t=0}^{T-1}\E\left[\left(\sum_{i=1}^N \omega_iU(z_{i,t})\right) - e_t\right]\\
&\quad\quad \textrm{s.t.}  & (\ref{p1xdcst}), (\ref{p1xucst}), (\ref{p1dtcst}), (\ref{p1utcst}),  (\ref{p1ccst}),
(\ref{p2aux0}), \textrm{ and }(\ref{p2aux}).
\end{align*}

Compared with \textbf{P1}, \textbf{P2} is over
  $\mb{x}_{d,t}$, $\mb{x}_{u,t}$ and $\mb{z}_{t}$ with  two more constraints (\ref{p2aux0}) and (\ref{p2aux}). Nevertheless,  \textbf{P2} contains no function of time average; instead, it maximizes  the long-term  time average of  the expected social welfare.

Denote $(\mb{x}_{d,t}^{\textrm{opt}}, \mb{x}_{u,t}^{\textrm{opt}})$ as an optimal solution to \textbf{P1}, and $(\mb{x}_{d,t}^{*},\mb{x}_{u,t}^{*}, \mb{z}_{t}^{*})$ as an optimal solution to \textbf{P2}. Define $\bar{\mb{z}}_{t}^{\textrm{opt}}\define [\bar{z}_{1,t}^{\textrm{opt}},\cdots,\bar{z}_{N,t}^{\textrm{opt}}]$ with the $i$-th element
\begin{align*}
\bar{z}_{i,t}^{\textrm{opt}} \define \lim_{T\to\infty} \frac{1}{T}\sum_{\tau=0}^{T-1} \E[x_{i,\tau}^{\textrm{opt}}], \quad \forall i, t,
\end{align*}
where $x_{i,\tau}^{\textrm{opt}} \define \1_{d,\tau}x_{id,\tau}^{\textrm{opt}} +\1_{u,\tau}x_{iu,\tau}^{\textrm{opt}}$.
Denote  the objective functions of \textbf{P1} and \textbf{P2}  as $f_1(\cdot)$ and $f_2(\cdot)$, respectively. The equivalence of \textbf{P1} and \textbf{P2} is stated below.
\begin{lemma}\label{lem-p1p2}
\textbf{P1} and \textbf{P2} have the same optimal objective, \ie, $f_1(\mb{x}_{d,t}^{\textrm{opt}}, \mb{x}_{u,t}^{\textrm{opt}}) = f_2(\mb{x}_{d,t}^{*},\mb{x}_{u,t}^{*}, \mb{z}_{t}^{*})$.  Furthermore, $(\mb{x}_{d,t}^{\textrm{opt}}, \mb{x}_{u,t}^{\textrm{opt}}, \bar{\mb{z}}_{t}^{\textrm{opt}})$ is an optimal solution to \textbf{P2}, and $(\mb{x}_{d,t}^{*},\mb{x}_{u,t}^{*})$ is an optimal solution to \textbf{P1}.
\end{lemma}
\IEEEproof
The proof follows the general framework given in \cite{bkneely}.
Details specific to our system are given in Appendix \ref{app-lemp1p2}.
\endIEEEproof

Lemma \ref{lem-p1p2} indicates that the transformation from  \textbf{P1} to \textbf{P2} results in no loss of optimality. Thus, in the following, we will focus on solving \textbf{P2} instead.

\vspace{-0.3cm}
\subsection{Problem Relaxation}\label{subsec-prorel}
 \textbf{P2} is still a challenging problem  since in the constraints (\ref{p1xdcst}) and (\ref{p1xucst}), the regulation allocation amount of each EV  depends on its current  energy state $s_{i,t}$, hence coupling with all previous regulation allocation amounts. To avoid such coupling, we
relax the constraints of $x_{id,t}$ and $x_{iu,t}$, and introduce the optimization problem \textbf{P3} below. \\
{\bf P3: }
\vspace{-0.3cm}
\begin{align}
\nonumber
\max_{\mb{x}_{d,t},\mb{x}_{u,t},\mb{z}_{t}} &\quad
\lim_{T\to\infty} \frac{1}{T}\sum_{t=0}^{T-1}\E\left[\left(\sum_{i=1}^N \omega_iU(z_{i,t})\right) - e_t\right]\\
\label{p3xd}
\textrm{s.t.} \quad & 0\le x_{id,t}\le \1_{i,t}x_{i,\max}, \forall i, t,\\
\label{p3xu}
&0\le x_{iu,t}\le \1_{i,t}x_{i,\max}, \forall i, t,\\
\label{p3xdu}
& \lta\E[b_{i,t}]=0,\forall i, \\
\nonumber
& (\ref{p1dtcst}), (\ref{p1utcst}),  (\ref{p1ccst}),
(\ref{p2aux0}), \textrm{ and }(\ref{p2aux}),
\end{align}
where in \eqref{p3xdu} $b_{i,t}$ is the effective charging/discharging amount defined in \eqref{bit}.
In \textbf{P3}, we have replaced the  constraints (\ref{p1xdcst}) and (\ref{p1xucst}) in \textbf{P2} with \eqref{p3xd}--\eqref{p3xdu}, thus have  removed the dependence of the regulation amount on $s_{i,t}$. We next
demonstrate that, any
 $(\mb{x}_{d,t},\mb{x}_{u,t})$ that satisfies (\ref{p1xdcst}) and (\ref{p1xucst}) also satisfies \eqref{p3xd}--\eqref{p3xdu}. Therefore,
\textbf{P3} is  a relaxed problem of  \textbf{P2}.

Consider the $i$-th EV. The constraints   (\ref{p1xdcst}) and (\ref{p1xucst}) in \textbf{P2} are equivalent to the following two sub-constraints:\\
if $\1_{i,t} = 1$, then
\begin{align}\label{xidm}
0\le x_{id,t}\le x_{i,\max}\\
\label{xium}
0\le x_{iu,t}\le x_{i,\max}\\
\label{st1}
s_{i,\min}\le s_{i,t+1}\le s_{i,\max};
\end{align}
if $\1_{i,t} = 0$, then
\begin{align}
\label{xidu0}
x_{id,t} = x_{iu,t} = 0.
\end{align}
Since
\eqref{xidm}, \eqref{xium}, and \eqref{xidu0} are equivalent to  \eqref{p3xd} and \eqref{p3xu},  we are left to justify  that \eqref{st1} (\ie, the boundedness of $s_{i,t}$)  implies \eqref{p3xdu}.  Recall that $s_{i,t}$ is  bounded for any returning time slot $t\in \mc{T}_{i,r}$ by the EV's self-control. Together, we need to justify that if
  $s_{i,t}\in [s_{i,\min}, s_{i,\max}], \forall t\in \mc{T}_{i,p}\cup\mc{T}_{i,l}$, then the constraint \eqref{p3xdu} holds. This result is shown in the following lemma.

\begin{lemma}\label{lem-cst3}
For the $i$-th EV,
 under the assumption A2,
 if
  $s_{i,t}\in [s_{i,\min}, s_{i,\max}], \forall t\in \mc{T}_{i,p}\cup\mc{T}_{i,l}$, then the constraint \eqref{p3xdu} holds, \ie,  $\lta\E[b_{i,t}]=0$.
\end{lemma}
 \IEEEproof
 See Appendix \ref{app-lemcst3}.
 \endIEEEproof

 From Lemma \ref{lem-cst3}, we know that,  the boundedness of $s_{i,t}$ indeed  implies \eqref{p3xdu}, which completes our demonstration that \textbf{P3} is a relaxed version of
  \textbf{P2} with a larger feasible solution set.  Later, we will  show in Section \ref{subsec:prop} that our proposed
algorithm for \textbf{P3} in fact ensures the boundedness of $s_{i,t}$, and thus provides a feasible solution to \textbf{P2} and to the original problem \textbf{P1}.

The relaxed problem \textbf{P3}     allows us to apply Lyapunov
optimization  to design a real-time algorithm for solving
welfare maximization.
To our best knowledge,
 this
 relaxation technique to accommodate  the type of time-coupled
 action constraints such as  (\ref{p1xdcst}) and (\ref{p1xucst}) is
first introduced in \cite{rah11} for a power-cost minimization problem
in data centers equipped with an energy storage device.  Unlike in \cite{rah11}, the structure of our problem is
more complicated, where the dynamics of the distributed storage devices (EVs) are considered, as well as a nonlinear objective which allows both
positive and negative values for the energy requirement $G_t$. Thus, the algorithm design is more
involved to ensure that the original constraints in \textbf{P2} are satisfied.

\vspace{-0.2cm}
\subsection{WMRA Algorithm}\label{subsec-wmar}
In this subsection,
we  propose a WMRA algorithm to  solve \textbf{P3} by   employing
 Lyapunov optimization technique.

We first define three virtual queues for each EV with the associated queue backlogs  $J_{i,t}$,  $H_{i,t}$,  and $K_{i,t}$.  The evolutionary behaviors of  $J_{i,t}$, $H_{i,t}$, and $K_{i,t}$ are as follows:
\begin{align}\label{jq}
&\hspace{-0.2cm}J_{i,t+1} = [J_{i,t} + \1_{d,t}C_i(x_{id,t}) + \1_{u,t}C_i(x_{iu,t})-c_{i,\textrm{up}}]^+;\\
\label{hq}
&\hspace{-0.2cm}H_{i,t+1} = H_{i,t} + z_{i,t}-x_{i,t};
\end{align}
\begin{subnumcases}{\label{kq} K_{i,t} =}
s_{i,t} - c_i, & if  $t\in \mc{T}_{i,r}$ \label{kq1}\\
K_{i,t-1} +b_{i,t-1},& otherwise, \label{kq2}
\end{subnumcases}
where  in \eqref{kq1} we have designed the constant
 $c_i = s_{i,\min}+2x_{i,\max}+V(\omega_i\mu+ e_{\max})$ with
 $V\in (0,V_{\max}]$ and
\begin{align}\label{vmax}
V_{\max} = \min_{1\le i \le N}\Big\{\frac{s_{i,\max}-s_{i,\min}-4x_{i,\max}}{2(\omega_i\mu+ e_{\max})}\Big\}.
\end{align}

The role of $V$ will be explained later. It will  also be clear in Section \ref{subsec:prop} that the specific expressions of $c_i$ and $V_{\max}$ are designed to ensure the boundedness of  $s_{i,t}$.
Note that
$x_{i,\max}$ is generally much smaller than the energy capacity. For example, for  the Tesla Model S base model\cite{tes}, the energy capacity  is $40$ kWh, and $x_{i,\max}=0.166$ kWh if the maximum charging rate $10$ kW is applied and the regulation duration is $1$ minute. Therefore,  $V_{\max}>0$ holds in general.

 From \eqref{kq1}, $K_{i,t}$ is re-initialized as a shifted version of $s_{i,t}$ every time the $i$-th EV returning to the aggregator-EVs system;
also, from \eqref{kq2}, $K_{i,t}$ evolves the same as $s_{i,t}$ for $t\in  \mc{T}_{i,p}$ (recall that the dynamics of $s_{i,t}$ may not follow \eqref{sev} when $\mb{1}_{i,t}=0$). Therefore, $K_{i,t}$ is essentially a shifted version of $s_{i,t}, \forall  t\in  \mc{T}_{i,p}\cup \mc{T}_{i,l}$, \ie,
\begin{align}\label{ksft}
K_{i,t} = s_{i,t}-c_i, \quad  \forall  t\in  \mc{T}_{i,p}\cup \mc{T}_{i,l}.
\end{align}
Additionally,
since the effective charging/discharging amount $b_{i,t} = 0$ when $\1_{i,t} = 0$, once the $i$-th EV leaves the system, the value of $K_{i,t}$ will be locked  until the next returning time slot of the EV, \ie,
\begin{align*}
&K_{i,t} = K_{i,t_{il,k}},  \quad \forall t\in \{t_{il,k},\cdots,t_{ir,k+1}-1\}, \\
&\hspace{3cm}\textrm{ and }\forall k\in\{1,2,\cdots\}.
\end{align*}

By introducing the virtual queues,  the constraints (\ref{p1ccst}) and
 (\ref{p2aux}) hold if the  queues $J_{i,t}$ and $H_{i,t}$
are mean rate stable, respectively\cite{bkneely}.  Below we give the definition of mean rate stability of a queue.

\emph{Definition:}
A queue $Q_t$ is mean rate stable if $\lim_{t\to \infty}\frac{\E[|Q_t|]}{t} =0$.

Unlike  $J_{i,t}$ and $H_{i,t}$,  since $K_{i,t}$ is re-initialized when $t\in \mc{T}_{i,r}$,  a new virtual queue is essentially created every time  the   $i$-th EV re-joining the system. Therefore, the
 mean rate stability of $K_{i,t}$ is  insufficient for the constraint \eqref{p3xdu} to hold, and a stronger condition is required.
 Fortunately, since  $K_{i,t}$ is just a shifted version of $s_{i,t}$  from \eqref{ksft},  based on  Lemma \ref{lem-cst3},  the following result is straightforward.
\begin{lemma}\label{lem-kcst}
For the $i$-th EV, under the assumption A2,
 if
  $K_{i,t}\in [s_{i,\min}-c_i, s_{i,\max}-c_i], \forall t\in \mc{T}_{i,p}\cup\mc{T}_{i,l}$, then the constraint \eqref{p3xdu} holds, \ie,  $\lta\E[b_{i,t}]=0$.
\end{lemma}

Later in Section \ref{subsec:prop}, we will show that by our proposed  algorithm the  boundedness assumption of $K_{i,t}$ in Lemma \ref{lem-kcst} can be guaranteed.

Define $\mb{J}_t \define [J_{1,t},\cdots, J_{N,t}]$, $\mb{H}_t \define [H_{1,t},\cdots, H_{N,t}]$,  $\mb{K}_t \define [K_{1,t},\cdots, K_{N,t}]$, and
$\mb{\Theta}_t \define [\mb{J}_t,\mb{H}_t, \mb{K}_t]$. Initialize $J_{i,0} = H_{i,0}= 0$, and $K_{i,0} = s_{i,0} - c_i, \forall i$.
Define the Lyapunov function $L(\mb{\Theta}_t ) \define \frac{1}{2}\sum_{i=1}^N(J_{i,t}^2+H_{i,t}^2+K_{i,t}^2),$ and the associated  one-slot Lyapunov drift as $\Delta(\mb{\Theta}_t ) \define \E\left[L(\mb{\Theta}_{t+1})-L(\mb{\Theta}_t ) |\mb{\Theta}_t \right].$ The drift-minus-welfare function is given by
$\Delta(\mb{\Theta}_t )-V\E\left[\sum_{i=1}^N \omega_iU(z_{i,t})-e_t|\mb{\Theta}_t\right],$
where $V\in (0, V_{\max}]$ is  the weight associated with the welfare objective. Hence,  the larger $V$, the more weight is put on the welfare objective.

Furthermore, we assume that for the $i$-th EV, the conditional expectation of the  energy state difference $\Delta_{i,k}$, given the queue backlogs before the EV returns,  is zero, \ie,
\begin{itemize}
\item[A3)] $\E[\Delta_{i,k}|\mb{\Theta}_t] = 0$, for $t = t_{ir,k+1}-1,\forall k\in\{1,2,\cdots\}, \forall i$.
\end{itemize}
Note that  A3 is mild, considering the random behavior of each EV due to other activities.

Now we provide an upper bound on the  drift-minus-welfare function in the following proposition.
\begin{pro}\label{pro-lydr}
Under the assumptions A1 and A3, the drift-minus-welfare function
 is upper-bounded as
 \setlength{\arraycolsep}{0pt}
\begin{align}
\nonumber
&\Delta(\mb{\Theta}_t )-V\E\left[\sum_{i=1}^N \omega_iU(z_{i,t})-e_t|\mb{\Theta}_t\right]\\
\nonumber
&\le B+ \sum_{i=1}^NK_{i,t}\E[b_{i,t}|\mb{\Theta}_t] + \sum_{i=1}^NH_{i,t}\E[z_{i,t}-x_{i,t}|\mb{\Theta}_t]\\
\nonumber
&+ \sum_{i=1}^NJ_{i,t}\E\left[\1_{d,t}C_{i}(x_{id,t})+\1_{u,t}C_i(x_{iu,t})-c_{i,\textrm{up}}|\mb{\Theta}_t \right]\\
\label{dpp}
&- V\E\left[\sum_{i=1}^N \omega_iU(z_{i,t})-e_t \Big|\mb{\Theta}_t\right],
\setlength{\arraycolsep}{0pt}
\end{align}
where
\begin{align}\label{bcst}
B\define \frac{1}{2}\sum_{i=1}^N\Big[2x_{i,\max}^2 +\Delta_{i,\max}^2 +[c_{i,\textrm{up}}^2, (c_{i,\max}-c_{i,\textrm{up}})^2]^+\Big].
\end{align}
\end{pro}
 \IEEEproof
See  Appendix \ref{app-prolydr}.
\endIEEEproof

Adopting the general framework of Lyapunov optimization\cite{bkneely},
we now propose the WMRA algorithm by
 minimizing the upper bound on the drift-minus-welfare function in  (\ref{dpp}) at each time slot. We will show in Section \ref{sec-perform} that the proposed algorithm can lead to a guaranteed performance.

 The minimization problem is
equivalent to  the following  decoupled  sub-problems  with respect to $\mb{z}_{t}$, $\mb{x}_{d,t}$, and $\mb{x}_{u,t}$, separately.
Denote the  solutions  produced by WMRA  as $\tilde{\mb{z}}_{t}\define[\tilde{z}_{1,t},\cdots,\tilde{z}_{N,t}]$, $\tilde{\mb{x}}_{d,t}\define[\tilde{x}_{1d,t},\cdots,\tilde{x}_{Nd,t}]$, and $\tilde{\mb{x}}_{u,t}\define[\tilde{x}_{1u,t},\cdots,\tilde{x}_{Nu,t}]$, respectively.
  Specifically,  we obtain $\tilde{z}_{i,t}, \forall i,$ by solving  \textbf{(a)}:
\begin{eqnarray*}\label{pa}
\textbf{(a): } \; \min_{z_{i,t}}\quad H_{i,t}z_{i,t} - \omega_iVU(z_{i,t}) \quad \textrm{s.t.}\;\; 0\le z_{i,t}\le x_{i,\max}.
\end{eqnarray*}
For $G_t>0$, we obtain $\tilde{\mb{x}}_{d,t}$ by   solving \textbf{(b1)}:
\begin{align}
\nonumber
\textbf{(b1): } &\min_{\mb{x}_{d,t}}\quad V e_{s,t}\big(G_t-\sum_{i=1}^Nx_{id,t}\big)  -\sum_{i=1}^NH_{i,t}x_{id,t}\\
\nonumber
&\quad\quad\quad +
\sum_{i=1}^N J_{i,t}C_i(x_{id,t})
+ \sum_{i=1}^NK_{i,t}x_{id,t}\\
\nonumber
 &\textrm{s.t.} \quad 0\le x_{id,t}\le \1_{i,t}x_{i,\max}, \quad \sum_{i=1}^Nx_{id,t}\le G_t.
\end{align}
For $G_t<0$, we obtain $\tilde{\mb{x}}_{u,t}$ by solving \textbf{(b2)}:
\begin{align}
\nonumber
\textbf{(b2): } &\min_{\mb{x}_{u,t}}\quad V e_{d,t}\big(|G_t|-\sum_{i=1}^Nx_{iu,t}\big)-\sum_{i=1}^NH_{i,t}x_{iu,t}\\
\nonumber
& \quad\quad\quad +
\sum_{i=1}^N J_{i,t}C_i(x_{iu,t})
- \sum_{i=1}^NK_{i,t}x_{iu,t} \\
\nonumber
 &\textrm{s.t.} \quad 0\le x_{iu,t}\le \1_{i,t}x_{i,\max}, \quad \sum_{i=1}^N x_{iu,t}\le |G_t|.
\end{align}
\begin{algorithm}[tb]
\caption{Welfare-maximizing  regulation allocation (WMRA) algorithm.}
\label{ag1}
\begin{algorithmic}[1]
\STATE The aggregator initializes the virtual queue vector $\mb{\Theta}_0$, and re-initialize  $K_{i,t} = s_{i,t}-c_{i}$ for $t\in \mc{T}_{i,r},  \forall i$.
\STATE At the beginning of each time slot $t$, the aggregator performs the following steps sequentially.

\begin{enumerate}[(2a)]
\item Observe  $G_t,  e_{s,t},  e_{d,t}, \1_t$,   $\mb{J}_{t}$,  $\mb{H}_{t}$, and $\mb{K}_t$.
\item Solve  (\textbf{a}) and record an optimal solution $\tilde{\mb{z}}_{t}$.  If $G_t>0$,
solve  (\textbf{b1}) and record an optimal solution $\tilde{\mb{x}}_{d,t}$. If $G_t<0$,
solve  (\textbf{b2}) and record an optimal solution $\tilde{\mb{x}}_{u,t}$. Allocate the regulation amounts among EVs based on $\tilde{\mb{x}}_{d,t}$ and $\tilde{\mb{x}}_{u,t}$.  If $\sum_{i=1}^N \tilde{x}_{id,t}<G_t$ or $\sum_{i=1}^N \tilde{x}_{iu,t}<|G_t|$,  clear the imbalance  using   external energy sources.
\item Update the virtual queues ${J}_{i,t},$  ${H}_{i,t},$ and $K_{i,t}, \forall i$, based on (\ref{jq}),  (\ref{hq}), and \eqref{kq2}, respectively.
\end{enumerate}
\vspace{-0.2cm}
\end{algorithmic}
\end{algorithm}

Note that \textbf{(a)}, \textbf{(b1)}, and \textbf{(b2)}
are all convex problems, so they can be efficiently solved using standard methods such as the interior point method\cite{bkboyd}.
We summarize  WMRA in Algorithm 1. Note from Steps (2b) and (2c) that, the solutions of \textbf{(a)} and \textbf{(b1)} (or \textbf{(b2)}) affect each other over multiple time slots through the update of   ${H}_{i,t}, \forall i$.
To perform  WMRA, no statistical information of the system is needed, which makes the algorithm easy to implement.

\vspace{-0.2cm}

\section{Performance Analysis}\label{sec-perform}
In this section,
we  characterize
the  performance of WMRA  with respect to our original problem \textbf{P1}.

\vspace{-0.2cm}
\subsection{Properties of WMRA Algorithm}\label{subsec:prop}
We now show that  WMRA  can ensure the boundedness of each EV's energy state.
The following lemma  characterizes  sufficient conditions under which the  solution of $\tilde{x}_{id,t}$ and $\tilde{x}_{iu,t}$ under WMRA  is zero.
\begin{lemma}\label{lem-xdu0}  Under the  WMRA algorithm, for any $t\in \mc{T}_{i,p}$,
\begin{enumerate}
\item for $G_t>0$, if $K_{i,t}> x_{i,\max} + V(\omega_i\mu+ e_{\max})$,  then $\tilde{x}_{id,t} = 0$, which means that $K_{i,t+1}$ cannot be increased at the next time slot; and
\item for $G_t<0$, if $K_{i,t}< -x_{i,\max} - V(\omega_i\mu+ e_{\max})$, then $\tilde{x}_{iu,t} = 0$, which means that $K_{i,t+1}$ cannot be decreased at the next time slot.
\end{enumerate}
\end{lemma}
\IEEEproof
See Appendix \ref{app-lemxdu0}.
\endIEEEproof

Since
Lemma  \ref{lem-xdu0} on the other hand provides conditions under which  queue  backlog $K_{i,t}$ can no longer increase or decrease,
using Lemma \ref{lem-xdu0}, we can prove the boundedness of    $K_{i,t}$ below.
\begin{lemma}\label{lem-kqbd}
Under the WMRA algorithm,  queue backlog $K_{i,t}$ associated with the $i$-th EV is bounded within $[s_{i,\min}-c_i, s_{i,\max}-c_i], \forall t\in\mc{T}_{i,p}\cup\mc{T}_{i,l}$.
\end{lemma}
\IEEEproof
See Appendix \ref{app-lemkqbd}.
\endIEEEproof
In the proof of Lemma \ref{lem-kqbd}, we remark on the specific designs of $c_i$ and $V_{\max}$, which are to ensure the boundedness of $K_{i,t}$ within a shifted preferred energy range.

From Lemma \ref{lem-kqbd}, the boundedness condition  of $K_{i,t}$ in  Lemma \ref{lem-kcst} is now satisfied,  therefore the conclusion there is true under WMRA.
Since $K_{i,t} = s_{i,t}-c_i, \forall t\in \mc{T}_{i,p} \cup \mc{T}_{i,l}$, using Lemma \ref{lem-kqbd}, the following lemma is straightforward.
\begin{lemma}\label{lem-sbd}
Under the WMRA algorithm,  the energy state of  the $i$-th EV is bounded within $[s_{i,\min},  s_{i,\max}], \forall t\in\mc{T}_{i,p}\cup\mc{T}_{i,l}$.
\end{lemma}

From Lemma \ref{lem-sbd},  the constraints (\ref{p1xdcst}) and (\ref{p1xucst}) in \textbf{P2} are met under WMRA.

\vspace{-0.3cm}
\subsection{Optimality of WMRA Algorithm}
In this subsection, we
investigate the optimality of WMRA
 by considering EVs with  both predictable and random dynamics, which are described below.
\begin{enumerate}
\item
 EVs with predictable dynamics:
Predictable dynamics could happen when each EV joins and leaves the aggregator-EVs system regularly (e.g. from $9$am to $12$pm in the morning, then from $2$pm to $6$pm in the afternoon).  Therefore, the   leaving and returning time slots of each EV can be predicted by the aggregator. In other words,
 the aggregator is aware of the realization of  $\1_{t},\forall t$ in advance. In this case, the  random system state at time slot $t$ is defined as $A_t \define (G_t, e_{s,t}, e_{d,t})$.  A specific case of  EVs with predictable dynamics is  static EVs, \ie, $\1_{i,t} = 1, \forall i, t$\footnote{In this work, we focus on the investigation of  non-static EVs. For static EVs,  the interested reader is   referred to  \cite{sun13} for details.}.

\item EVs with random dynamics:   If the EVs do not participate in  the aggregator-EVs system regularly, then
the aggregator cannot predict their dynamics beforehand, and therefore has to observe $\1_t$ every time slot. In this case, the random  system state at time slot $t$ is defined as $A_t \define (G_t, e_{s,t}, e_{d,t},\1_t)$.
\end{enumerate}
Note that the WMRA algorithm is the same under both of the above  cases. The only difference between them is that, in the optimization problem \textbf{P3}, the expectations are taken over different randomness of the system state.
The performance under WMRA as compared to the optimal solution of \textbf{P1} is given in
 the following theorem, which  applies to both   predictable and random dynamics.
\begin{theorem}\label{the-iid}
Under the assumptions A1, A2, and A3, given  the system state  $A_t$  is i.i.d. over time,
\begin{enumerate}
\item $(\tilde{\mb{x}}_{d,t}, \tilde{\mb{x}}_{u,t})$ is feasible for \textbf{P1}, \ie, it satisfies \eqref{p1xdcst}--\eqref{p1ccst}.
\item
$
f_1(\tilde{\mb{x}}_{d,t}, \tilde{\mb{x}}_{u,t}) \ge f_1({\mb{x}}^{\textrm{opt}}_{d,t}, {\mb{x}}^{\textrm{opt}}_{u,t})-\frac{B}{V},
$
where  $B$ is defined in \eqref{bcst}  and $V\in (0, V_{\max}]$.
\end{enumerate}
\end{theorem}
\IEEEproof
See Appendix \ref{app-theiid}.
\endIEEEproof
\emph{Remarks:}
From Theorem \ref{the-iid}, the welfare performance of WMRA  is away from the optimum by $O(1/V)$.
Hence, the larger $V$, the better the performance of WMRA. However, in practice, due to the  boundedness condition of EV's battery capacity, $V$ cannot be arbitrarily large and is upper bounded by $V_{\max}$, which is defined in \eqref{vmax}.  Note that  $V_{\max}$ increases with the smallest span of the EVs' preferred battery capacity ranges, \ie, $\min_{1\le i\le N}\{s_{i,\max}-s_{i,\min}\}$. Therefore, roughly speaking, the performance gap between  WMRA and the optimum  decreases as the smallest battery capacity   increases.  Asymptotically, as the  EVs' battery capacities go to infinity, WMRA would  achieve exactly the optimum.

In  Theorem \ref{the-iid}, the i.i.d. condition of $A_t$ can be relaxed to  Markovian, and a similar performance bound can be obtained.  In particular, this relaxed condition can accommodate the case  where $G_t$ is Markovian and has a ramp rate constraint ($|G_t-G_{t-1}| \le $ ramp rate $\times \Delta t$), by properly designing the transition probability matrix of $G_t$.
\begin{theorem}\label{the-mak}
Under the assumptions A1, A2, and A3, given that  the system state  $A_t$ evolves based on a finite state irreducible and aperiodic Markov chain,\begin{enumerate}
\item $(\tilde{\mb{x}}_{d,t}, \tilde{\mb{x}}_{u,t})$ is feasible for \textbf{P1}, \ie, it satisfies \eqref{p1xdcst}--\eqref{p1ccst}.
\item
$
f_1(\tilde{\mb{x}}_{d,t}, \tilde{\mb{x}}_{u,t}) \ge f_1({\mb{x}}^{\textrm{opt}}_{d,t}, {\mb{x}}^{\textrm{opt}}_{u,t})-O(1/V),
$
where
$V\in (0, V_{\max}]$.
\end{enumerate}
\end{theorem}
\IEEEproof
The above results can be proved
  by expanding the proof of Theorem \ref{the-iid}  using a   multi-slot drift technique \cite{bkneely}. We  omit the proof here for brevity.
\endIEEEproof

\vspace{-0.2cm}

\section{Simulation Results}\label{sec-sim}
Besides the analytical performance bound derived above,
we are further interested in evaluating  WMRA in example numerical settings. Towards this goal, we have developed an aggregator-EVs model in Matlab and compared WMRA with a greedy algorithm.

Suppose that the aggregator is connected with $N=100$ EVs, evenly split into Type I  (based on  the 2012 Ford Focus Electric)  and Type II  (based on the Tesla Model S base model).
The parameters of Type I and Type II EVs are summarized in Table I \cite{tes,for}. The  maximum regulation amount $x_{i,\max}$ can be derived by multiplying the maximum charging/discharging rate with the regulation interval $\Delta t$.
In current practice, $\Delta t$ is of the order of seconds. For example, for PJM, $\Delta t =2$ seconds \cite{pjm}, and for NYISO, $\Delta t =6$ seconds \cite{nyiso}.  In simulations, we set
$\Delta t= 5$ seconds as an example.

\begin{figure}[t]
\begin{center}
\includegraphics[height=0.9in,width=2.2in]{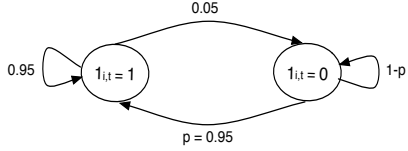}
\end{center}
\vspace{-0.4cm}
\caption{Transition probabilities of $\1_{i,t},\forall i$.}
\label{fig-mc}
\vspace{-0.4cm}
\end{figure}

Consider that the system state $A_t = (G_t, e_{s,t}, e_{d,t},\1_t)$ follows a finite state irreducible and aperiodic Markov chain. For the regulation signal $G_t$, we ignore the ramp rate constraint in our simulations. At each time slot, we draw a sample of $G_t$  from  a uniformly distributed set $\{-1.15,-1.15+\Delta_1,-1.15+2\Delta_1,\cdots,1.15\}$ (kWh) with the cardinality $200$, where  $1.15$ kWh is   the  maximum  allowed regulation  amount  at each time slot if all $N$ EVs are in the system.
The unit costs of the external sources,  $e_{s,t}$ and $e_{d,t}$, are drawn  uniformly from  a discrete set $\{0.1, 0.1+\Delta_2, 0.1+2\Delta_2,\cdots, 0.12\}$ (dollars/kWh) with the cardinality $200$. The lower bound $0.1$ dollars/kWh and the upper bound $0.12$ dollars/kWh correspond to the mid-peak and the on-peak electricity prices in Ontario, respectively \cite{onep}.
The dynamics of each EV is described by the indicator random variable $\1_{i,t}$, which represents whether the $i$-th EV is in the system at time slot $t$. In particular, we assume that $\1_{i,t}$ follows a two-state Markov chain as shown in Fig.~\ref{fig-mc}.
The state transition probability $p \define \p(0\to 1)$ is set to be $0.95$ by default.

For the $i$-th EV,
 the $(k+1)$-th returning energy state  $s_{i,t_{ir,k+1}}$  is drawn uniformly from the interval $[s_{i,t_{il,k}}-\Delta_3, s_{i,t_{il,k}}+\Delta_3]$, where $s_{i,t_{il,k}}$ is the $k$-th leaving energy state of the $i$-th EV and $\Delta_3 = 5\%s_{i,\textrm{cap}}$\footnote{We ensure that all returning energy states are  within the preferred range $[s_{i,\min}, s_{i,\max}]$ by ignoring  unqualified samples.}.
We set the minimum preferred energy state $s_{i,\min} = 0.1s_{i,\textrm{cap}}$,  and the maximum preferred energy state $s_{i,\max} = 0.9s_{i,\textrm{cap}}$ except otherwise mentioned.
In the  objective function of \textbf{P1}, we set $U(x) = \log(1+x)$ and $\omega_i = 1,\forall i$.
Since the degradation cost function $C_i(\cdot)$ is proprietary  and unavailable, in simulations, we set  $C_i(x) = x^2$ as an example. The upper bound $c_{i,\textrm{up}}$ is set to be ${x_{i,\max}^2}/{4}$.

\begin{table}[t]\label{typ12}
\renewcommand{\arraystretch}{1.2}
\caption{Parameters of Type I and Type II EVs }
\label{flopadapR3}
\centering
\begin{tabular}{p{4.8cm}|p{1.2cm}|p{1.4cm}}
\hline
 & Type I EV & Type II EV\\
\hline
Energy capacity $s_{i,\textrm{cap}}$ (kWh) & 23 & 40\\
\hline
Maximum charging/discharging rate (kW) & $ 6.6$ & $ 10$\\
\hline
\end{tabular}
\end{table}

\begin{figure}[t]
\begin{center}
\includegraphics[height=2.8in,width=3.2in]{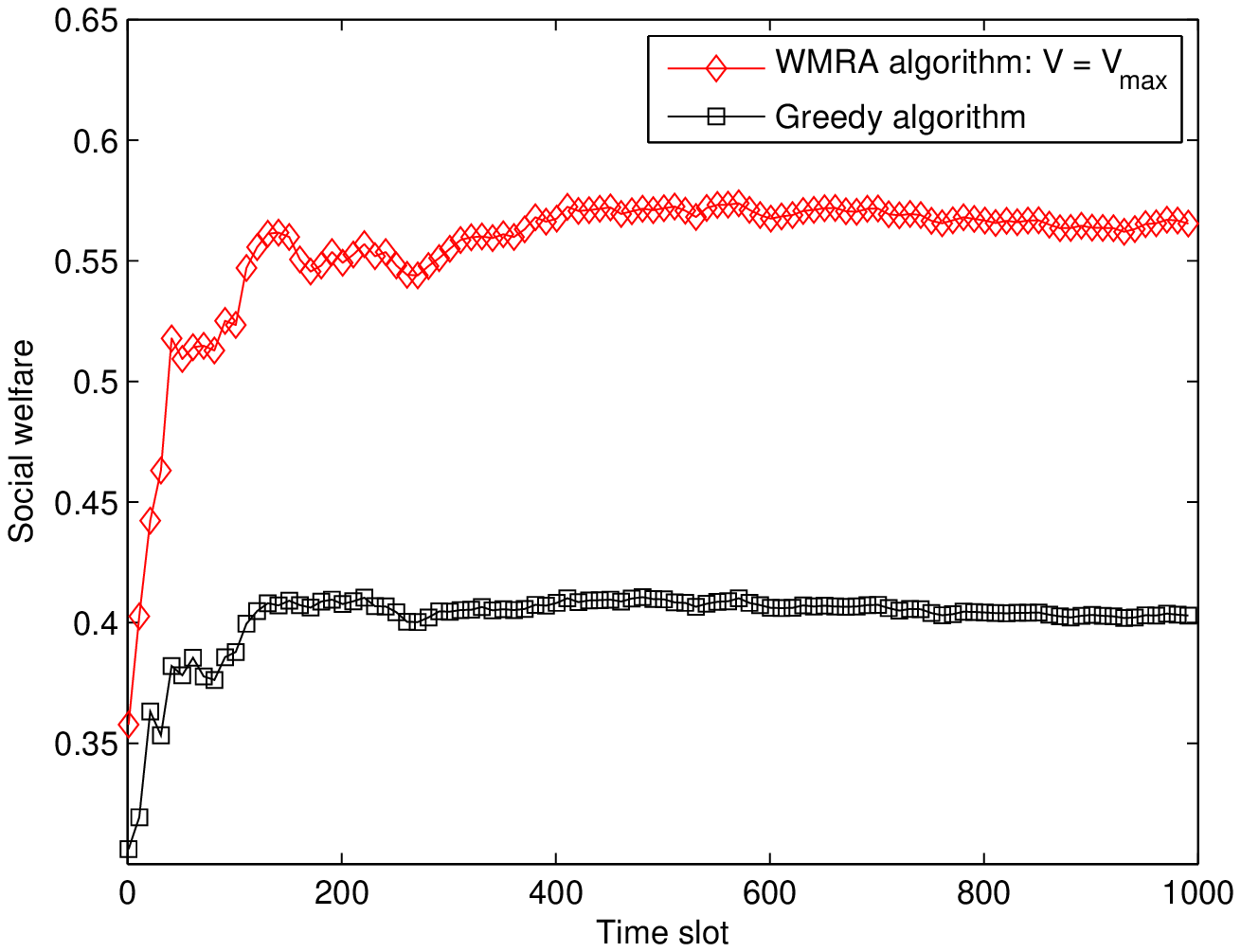}
\end{center}
\vspace{-0.4cm}
\caption{Time-averaged social welfare with  $V=V_{\max}$.}
\label{fig-vmax}
 \begin{center}
  \includegraphics[height=2.8in,width=3.2in]{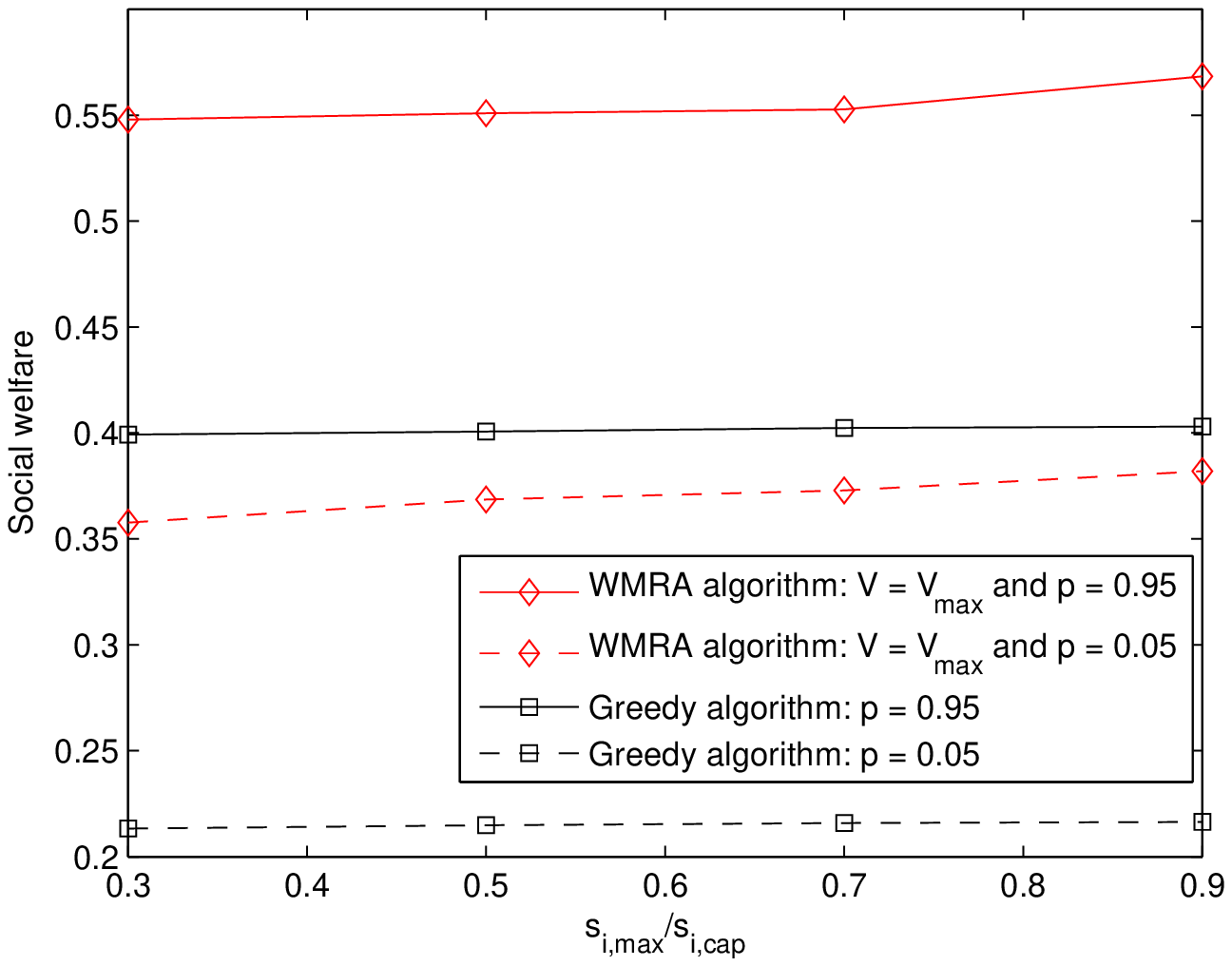}
 \end{center}
 \vspace{-0.4cm}
 \caption{Time-averaged social welfare with  various $s_{i,\max}$ and $V=V_{\max}$.}
 \label{fig-smax}
 \vspace{-0.2cm}
 \end{figure}

To allocate the requested regulation amount, we apply WMRA  in Algorithm 1 at each time slot.
  For comparison, we consider a greedy algorithm which only optimizes the system performance at the current time slot.  The regulation allocation  at each time slot is determined by  the following  optimization problem.
\begin{align*}
&\max_{\mb{x}_{d,t},\mb{x}_{u,t}}\quad
\left(\sum_{i=1}^N \omega_iU(x_{i,t})\right) - e_t\\
&\quad \textrm{s.t.} \quad \quad (\ref{p1xdcst}), (\ref{p1xucst}),  \eqref{p1dtcst}, \eqref{p1utcst}, \textrm{ and }\\
& \quad\quad\quad\quad
\1_{d,t}C_i(x_{id,t}) + \1_{u,t}C_i(x_{iu,t})\le c_{i,\textrm{up}},\forall i.
\vspace{-0.1cm}
\end{align*}
The above problem
  is a convex optimization problem,  and we use the standard solver in MATLAB to obtain its solution.

In Figs. \ref{fig-vmax} and \ref{fig-smax}, we compare the performance of WMRA with $V = V_{\max}$ and the performance of the greedy algorithm.  From  Fig. \ref{fig-vmax}, with $s_{i,\max} = 0.9s_{i,\textrm{cap}}$, WMRA is uniformly superior to the greedy algorithm over all time slots, with the advantage about $40\%$.
In Fig. \ref{fig-smax},  we set the transition probability $p$ to be $0.95$ and $0.05$, and  vary $s_{i,\max}$ from $0.3s_{i,\textrm{cap}}$ to  $0.9s_{i,\textrm{cap}}$. For $p = 0.95$, the observations are as follows. First, WMRA uniformly outperforms the greedy algorithm over different values of
$s_{i,\max}$.
 Second, as $s_{i,\max}$ increases, the social welfare under WMRA slightly rises. This is because increasing $s_{i,\max}$ effectively increases $V_{\max}$, which improves the performance  of WMRA. This observation is also
consistent with the remarks after Theorem  \ref{the-iid}.    In contrast,   the greedy algorithm cannot benefit from the expanded energy range.
For $p = 0.05$, the trends of the curves resemble those for $p=0.95$, but the social welfare of both algorithms drops. This is because when $p$ is decreased,
roughly speaking, there  are fewer EVs in the system for the regulation service.  Hence, to provide the requested regulation amount, the aggregator more relies on the expensive external energy sources, which  leads to a decreased social welfare.

  \begin{figure}[t]
 \begin{center}
 \includegraphics[height=2.8in,width=3.2in]{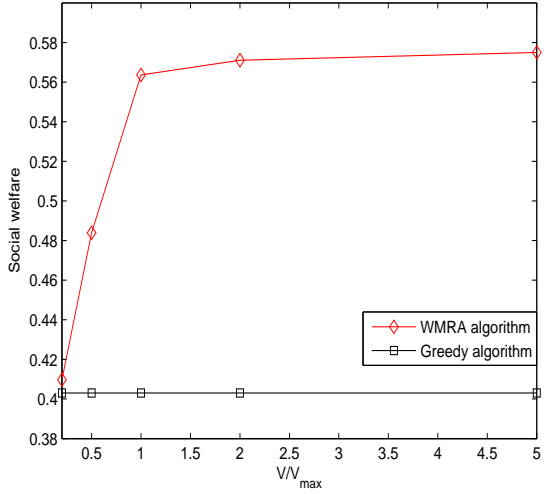}
 \end{center}
 \vspace{-0.4cm}
 \caption{Time-averaged social welfare with   various values of $V$.}
 \label{fig-difv}
 \end{figure}

In Fig. \ref{fig-difv}, we show the performance of WMRA with the value of $V$  ranging from $0.2V_{\max}$ to $5V_{\max}$, and compare it with the performance of the greedy algorithm.
For WMRA, as expected, the social welfare grows with the value of $V$; also,  the growing rate slows down when $V$ gets larger.
Moreover, we observe that WMRA outperforms the greedy algorithm  even with $V = 0.2V_{\max}$.

\begin{figure}[t]
 \begin{center}
 \includegraphics[height=2.8in,width=3.2in]{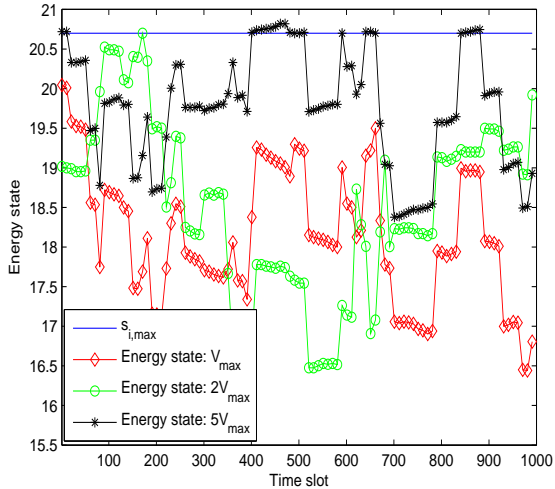}
 \end{center}
 \vspace{-0.2cm}
 \caption{Sample path of a Type I EV's energy state  with  $V= [1, 2, 5]V_{\max}$.}
 \label{fig-difvenergy}
 \end{figure}

In Lemma \ref{lem-sbd},  the energy state of each EV  is shown to be restricted  within $[s_{i,\min},s_{i,\max}]$ when $V\in (0,V_{\max}]$. In Fig. \ref{fig-difvenergy}, for $V$ being $V_{\max}, 2V_{\max}$, and $5V_{\max}$, we show the evolution  of a Type I EV's energy state under WMRA. We see that, when $V = V_{\max}$, the energy state  is always within  the preferred range. In contrast, when  $V = 2V_{\max}$ or $5V_{\max}$, the associated energy state can exceed the preferred range from time to time. Furthermore, the larger $V$ the more frequently such violation happens. Therefore,
the observations in Figs. \ref{fig-difv} and   \ref{fig-difvenergy}  demonstrate the significance of $V_{\max}$ in achieving the maximum social welfare under WMRA considering the
 constraint of EV's preferred energy range.

\section{Conclusion}\label{sec-con}
We  studied a practical model of a dynamic aggregator-EVs system providing regulation service to a power grid. We formulated the  regulation allocation optimization as a long-term time-averaged social welfare maximization problem. Our formulation 
 accounts for random system dynamics, battery constraints,  the costs for   battery degradation and  external energy sources, and especially, the dynamics of EVs. Adopting a general Lyapunov optimization framework, we  developed a real-time WMRA algorithm for the aggregator to fairly allocate the regulation amount among EVs. The algorithm does not require any knowledge of the statistics of the system state. We were able to bound the performance of WMRA  to that under the optimal solution, and showed that the performance of WMRA is asymptotically optimal as EVs' battery capacities go to infinity. Simulation demonstrated that WMRA offers substantial performance gains over a greedy algorithm that maximizes per-slot social welfare objective.

\appendices
\section{Proof of Lemma $1$}\label{app-lemp1p2}
It is easy to see that $(\mb{x}_{d,t}^*, \mb{x}_{u,t}^*)$ is feasible for \textbf{P1}. To show that $(\mb{x}_{d,t}^{\textrm{opt}},\mb{x}_{u,t}^{\textrm{opt}}, {\bar{\mb{z}}_{t}}^{\textrm{opt}})$ is feasible for
\textbf{P2}, it suffices to show that $\bar{\mb{z}}_{t}^{\textrm{opt}}$ satisfies
(\ref{p2aux0}) and (\ref{p2aux}). Using the definition of $\bar{z}_{i,t}^{\textrm{opt}}$, (\ref{p2aux}) naturally holds. Also, since $x_{i,t}^{\textrm{opt}}$ lies in $[0,x_{i,\max}]$, which is a closed interval, (\ref{p2aux0}) holds.

We claim that
\begin{eqnarray}
\nonumber
f_1(\mb{x}_{d,t}^{\textrm{opt}},\mb{x}_{u,t}^{\textrm{opt}})&&= f_2(\mb{x}_{d,t}^{\textrm{opt}},\mb{x}_{u,t}^{\textrm{opt}}, \bar{\mb{z}}_{t}^{\textrm{opt}})\\
\nonumber
&&\le f_2(\mb{x}_{d,t}^*, \mb{x}_{u,t}^*, \mb{z}_{t}^*)\\
\nonumber
&&\le f_1(\mb{x}_{d,t}^*, \mb{x}_{u,t}^*)\\
\label{f1f2}
&&\le f_1(\mb{x}_{d,t}^{\textrm{opt}},\mb{x}_{u,t}^{\textrm{opt}}).
\end{eqnarray}
Using the definition of $\bar{z}_{i,t}^{\textrm{opt}}$ in $f_2(\cdot)$,
the first equality holds. The first and the third inequalities hold since $(\mb{x}_{d,t}^*, \mb{x}_{u,t}^*, \mb{z}_{t}^*)$ and $(\mb{x}_{d,t}^{\textrm{opt}},\mb{x}_{u,t}^{\textrm{opt}})$ are optimal for $f_2(\cdot)$ and $f_1(\cdot)$, respectively. The second inequality is derived using Jensen's inequality for concave functions. Since (\ref{f1f2}) is satisfied with equality, all inequalities in (\ref{f1f2}) turn into equalities, which indicates the equivalence of \textbf{P1} and \textbf{P2}.

\section{Proof of Lemma \ref{lem-cst3}}\label{app-lemcst3}
Let $T$ be large enough.
 For the $i$-th EV,
 decompose the total  effective charging/discharging amount within $T$ time slots as
\begin{eqnarray}\label{abt}
\sum_{t=0}^{T-1}b_{i,t} = \sum_{t=0}^{t_{il,k^*}-1}b_{i,t}  +  \sum_{t=t_{il,k^*}}^{T-1}b_{i,t},
\end{eqnarray}
where $k^*\define \max\{k: t_{il,k}\le (T-1), k\in\{1,2,\cdots\} \}$ is defined to be the total number of the leaving times of the $i$-th EV up to time slot $T-1$. On the right hand side of \eqref{abt}, the first term corresponds to the total effective charging/discharging amount  before the last leaving time, and the second term corresponds to the rest of the total effective charging/discharging  amount. Using the decomposition in \eqref{abt},
to show \eqref{p3xdu}, it suffices to show that the two limits $\lim_{T\to\infty}\frac{1}{T}\E[\sum_{t=0}^{t_{il,k^*}-1}b_{i,t}]$ and $\lim_{T\to\infty}\frac{1}{T}\E[\sum_{t=t_{il,k^*}}^{T-1}b_{i,t} ]$ are both equal to zero.

First consider the second limit. For the $i$-th EV, if there is no return  between $t_{il,k^*}$ and $T-1$, then $\sum_{t=t_{il,k^*}}^{T-1}b_{i,t} = 0$ and thus $\lim_{T\to\infty}\frac{1}{T}\E[\sum_{t=t_{il,k^*}}^{T-1}b_{i,t} ]=0$. Or, if there is one return, then  $\sum_{t=t_{il,k^*}}^{T-1}b_{i,t} = s_{i,T} - s_{i,t_{ir,k^*+1}}$. Using the boundedness condition of $s_{i,t}$, we have $\lim_{T\to\infty}\frac{1}{T}\E[\sum_{t=t_{il,k^*}}^{T-1}b_{i,t} ]=0$. Together, the second limit is zero.

Next we  show that the first limit is also zero. Based on the energy state evolution in  \eqref{sev}, there is
 \setlength{\arraycolsep}{0pt}
\begin{eqnarray}\nonumber
\sum_{t=0}^{t_{il,k^*}-1}b_{i,t}&&= \sum_{k=1}^{k^*}s_{i,t_{il,k}}  -\sum_{k=1}^{k^*}s_{i,t_{ir,k}} \\
\label{sdk1}
&&=s_{i,t_{il,k^*}}-s_{i,t_{ir,1}} - \sum_{k=1}^{k^*-1}\Delta_{i,k}.
\end{eqnarray}
 \setlength{\arraycolsep}{0pt}
 Taking expectations of both sides of \eqref{sdk1}, dividing them by $T$, then taking limits gives
 \begin{eqnarray}\nonumber
 \lim_{T\to\infty}\frac{1}{T}\E&&\Bigg[\sum_{t=0}^{t_{il,k^*}-1}b_{i,t}\Bigg]
 = \lim_{T\to\infty}\frac{1}{T}\E\Bigg[s_{i,t_{il,k^*}}-s_{i,t_{ir,1}}\Bigg] \\
 \nonumber
 &&-\lim_{T\to\infty}\frac{1}{T}\E\Bigg[\sum_{k=1}^{k^*-1}\Delta_{i,k} \Bigg] = 0,
 \end{eqnarray}
 where the last equality is derived by  the boundedness of $s_{i,t}$ and the assumption A2.
 This completes the proof.

\section{Proof of Proposition \ref{pro-lydr}}\label{app-prolydr}
Based on the definition of $L(\mb{\Theta}_t)$, the difference \begin{eqnarray}\nonumber
&&\hspace{-0.5cm}L(\mb{\Theta}_{t+1})-L(\mb{\Theta}_t)  \\
\label{ld}
 &&\hspace{-1cm}=\frac{1}{2}\sum_{i=1}^NH_{i,t+1}^2+J_{i,t+1}^2+K_{i,t+1}^2 -H_{i,t}^2-J_{i,t}^2-K^2_{i,t}.
\end{eqnarray}
In (\ref{ld}),
$H_{i,t+1}^2-H_{i,t}^2$ and $J_{i,t+1}^2-J_{i,t}^2$  can be upper bounded as follows.
\setlength{\arraycolsep}{0pt}
\begin{align}
\label{hdup}
&H_{i,t+1}^2-H_{i,t}^2\le  2H_{i,t}(z_{i,t}-x_{i,t}) + x_{i,\max}^2\\
\nonumber
&J_{i,t+1}^2-J_{i,t}^2
\le 2J_{i,t}[\1_{d,t}C_i(x_{id,t})+\1_{u,t}C_i(x_{iu,t})-c_{i,\up}]\\
\label{jdup}
&\hspace{3cm}+[c_{i,\textrm{up}}^2,(c_{i,\max}-c_{i,\textrm{up}})^2]^+ .
\end{align}
\setlength{\arraycolsep}{0pt}
Taking  conditional expectations for both sides in \eqref{hdup} and \eqref{jdup}, we have
\begin{align}
\label{hcon}
&\E[H_{i,t+1}^2-H_{i,t}^2|\mb{\Theta}_t]\le 2H_{i,t}\E[z_{i,t}-x_{i,t}|\mb{\Theta}_t] + x_{i,\max}^2\\
\nonumber
&\E[J_{i,t+1}^2-J_{i,t}^2|\mb{\Theta}_t]\le 2J_{i,t}\E[\1_{d,t}C_i(x_{id,t})+\1_{u,t}C_i(x_{iu,t})\\
\label{jcon}
&\hspace{2cm}-c_{i,\up}|\mb{\Theta}_t] + [c_{i,\textrm{up}}^2,(c_{i,\max}-c_{i,\textrm{up}})^2]^+.
\end{align}

Now consider $K_{i,t+1}^2-K_{i,t}^2$.
 When $\1_{i,t} =1$, we have
 $K_{i,t+1} = K_{i,t} +b_{i,t}$ and thus
\begin{align}\label{kdup1}
K_{i,t+1}^2-K_{i,t}^2 \le 2K_{i,t}b_{i,t}+x_{i,\max}^2.
\end{align}
When $\1_{i,t} = 0$, we have $b_{i,t} =0$ and there are two cases. First, for $t\in \{t_{il,k},t_{il,k}+1,\cdots,t_{ir,k+1}-2\}, \forall k\in \{1,2,\cdots\}$, there is $K_{i,t+1}=  K_{i,t}$. So, we can express
\begin{align}\label{kdup2}
K_{i,t+1}^2-K_{i,t}^2 = 2K_{i,t}b_{i,t}.
\end{align}
Second, for $t = t_{ir,k+1}-1,\forall k\in\{1,2,\cdots\}$, we have  $K_{i,t} = s_{i,t_{il,k}}-c_i$ and $K_{i,t+1} = K_{i,t} + \Delta_{i,k}$. Hence, by the assumption A1,
\begin{align}\label{kdup3}
K_{i,t+1}^2-K_{i,t}^2 \le 2K_{i,t}\Delta_{i,k}+\Delta_{i,\max}^2.
\end{align}
Using the assumption A3, from
\eqref{kdup1}, \eqref{kdup2}, and \eqref{kdup3}, we have
\begin{align}\label{kdup}
\E[K_{i,t+1}^2-K_{i,t}^2|\mb{\Theta}_t] \le 2K_{i,t}\E[b_{i,t}|\mb{\Theta}_t] + x_{i,\max}^2 + \Delta_{i,\max}^2.
\end{align}
Using the definition of $\Delta(\mb{\Theta}_t)$ and the upper bounds in \eqref{hcon}, \eqref{jcon}, and  \eqref{kdup}, we can derive the upper bound on the drift-minus-welfare function in  Proposition \ref{pro-lydr}.

\section{Proof of Lemma \ref{lem-xdu0}}\label{app-lemxdu0}
We need the following lemma.
\begin{lemma}\label{lem-h}
Under the WMRA algorithm,  queue backlog $H_{i,t}$ associated with the $i$-th EV is  upper bounded as follows:
\begin{align*}
H_{i,t}\le V\omega_i\mu+x_{i,\max}.
\end{align*}
\end{lemma}
\IEEEproof
This can be shown using a similar method as in \cite{bkneely}, and the technical condition \eqref{ucond} is needed.
\endIEEEproof

1) Consider $G_t>0$. Suppose that when $K_{i,t}> x_{i,\max} + V(\omega_i\mu+ e_{\max})$, one optimal solution under WMRA is $\tilde{\mb{x}}_{d,t}$ with $\tilde{x}_{id,t}>0$. Then we show that  we can find  another  solution with $\tilde{x}_{jd,t}, \forall j \neq i$ and $\tilde{x}_{id,t} =0$  resulting in a strictly smaller objective value, which is a contradiction.

Using the objective function of (\textbf{b1}), this is equivalent to showing that
\begin{align*}
&Ve_{s,t}\left(G_t-\sum_{j=1}^N\tilde{x}_{jd,t}\right)  -\sum_{j=1}^NH_{j,t}\tilde{x}_{jd,t}\\
&+
\sum_{j=1}^N J_{j,t}C_j(\tilde{x}_{jd,t})
+ \sum_{j=1}^NK_{j,t}\tilde{x}_{jd,t} \\
&> V e_{s,t}\left(G_t-\sum_{j=1}^N\tilde{x}_{jd,t} + \tilde{x}_{id,t}\right)  -\sum_{j\neq i}H_{j,t}\tilde{x}_{jd,t}\\
&+
\sum_{j \neq i}J_{j,t}C_j(\tilde{x}_{jd,t})
+ \sum_{j \neq i}K_{j,t}\tilde{x}_{jd,t},
\end{align*}
which is equivalent to
\begin{eqnarray}\label{heq2}
-H_{i,t}\tilde{x}_{id,t} + J_{i,t}C_i(\tilde{x}_{id,t}) + K_{i,t}\tilde{x}_{id,t}  > V e_{s,t}\tilde{x}_{id,t}.
\end{eqnarray}
Since $J_iC_i(\tilde{x}_{id,t})\ge 0$, from \eqref{heq2}, it suffices to show that
\begin{eqnarray}\label{kxid}
(K_{i,t}-H_{i,t}-V e_{s,t})\tilde{x}_{id,t}>0.
\end{eqnarray}
Since $\tilde{x}_{id,t}>0$,
\eqref{kxid} is true by using the assumption that $K_{i,t}> x_{i,\max} + V(\omega_i\mu+ e_{\max})$  and Lemma \ref{lem-h} in which $H_{i,t}$ is  upper bounded.

2) Consider $G_t<0$. Suppose that when $K_{i,t}< -x_{i,\max} - V(\omega_i\mu+ e_{\max})$, one optimal solution under WMRA is $\tilde{\mb{x}}_{u,t}$ with $\tilde{x}_{iu,t}>0$. Then there is a contradiction since  we can construct another  solution with $\tilde{x}_{ju,t}, \forall j \neq i$ and $\tilde{x}_{iu,t} =0$ which results in a strictly smaller objective value. The proof is similar as  that in 1) and  is omitted here.

\section{Proof of Lemma \ref{lem-kqbd}}\label{app-lemkqbd}
Consider the   set $\{t_{ir,k},t_{ir,k}+1,\cdots, t_{il,k}\}$ for any $k\in \{1,2,\cdots\}$.
We show below that $K_{i,t}$ is bounded for any $t$ in such set by induction.

First consider the upper bound. For the time slot $t_{ir,k}$, based on \eqref{kq} and $s_{i,t_{ir,k}}\le s_{i,\max}$, there is $K_{i,t_{ir,k}}\le s_{i,\max}-c_i$.  Assume that the upper bound holds for time slot $t$ and  consider the following two cases of $K_{i,t}$.

Case $1$:
$x_{i,\max} + V(\omega_i\mu+ e_{\max}) < K_{i,t}\le s_{i,\max}-c_i$ (We can check that $x_{i,\max} + V(\omega_i\mu+ e_{\max})<s_{i,\max}-c_i$ since $V\le V_{\max}$). For $G_t>0$, from Lemma \ref{lem-xdu0} 1), there is $\tilde{x}_{id,t} = 0$. Therefore,  $K_{i,t+1} = K_{i,t}\le s_{i,\max}-c_i$.
For $G_t<0$, we have
$K_{i,t+1} = K_{i,t}-x_{iu,t}\le K_{i,t}\le s_{i,\max}-c_i.$

Case $2$: $K_{i,t}\le x_{i,\max} + V(\omega_i\mu+ e_{\max})$. From \eqref{kq}, $K_{i,t+1}\le 2x_{i,\max} + V(\omega_i\mu+ e_{\max})\le s_{i,\max}-c_i,$ where the last inequality holds since  $V\le V_{\max}$.

Now look at the lower bound. For the time slot $t_{ir,k}$,  based on \eqref{kq} and $s_{i,t_{ir,k}}\ge s_{i,\min}$, there is $K_{i,t_{ir,k}}\ge s_{i,\min}-c_i$. Assume that the lower bound holds for time slot $t$ and consider the following two cases of $K_{i,t}$.

 Case $1'$:  $s_{i,\min}-c_i\le K_{i,t}<-x_{i,\max} - V(\omega_i\mu+ e_{\max})$ (We can check that $s_{i,\min}-c_i<-x_{i,\max} - V(\omega_i\mu+ e_{\max})$ since $x_{i,\max}>0$). For $G_t<0$, from Lemma  \ref{lem-xdu0} 2), there is $\tilde{x}_{iu,t} = 0$. Therefore,   $K_{i,t+1} = K_{i,t} \ge s_{i,\min}-c_i,$
 For $G_t>0$, we have
$K_{i,t+1} = K_{i,t}+x_{id,t}\ge K_{i,t}\ge s_{i,\min}-c_i.$

Case $2'$: $K_{i,t}\ge -x_{i,\max} - V(\omega_i\mu+ e_{\max})$. From \eqref{kq}, $K_{i,t+1}\ge -2x_{i,\max}- V(\omega_i\mu+ e_{\max})$, which is exactly $s_{i,\min}-c_i$.

\emph{Remarks:}
To track the energy state $s_{i,t}$, in principle, the shift $c_{i}$ can be any  number. However,  to make the proof in Case $2'$ work, $c_i$ is lower bounded, \ie, should satisfy  $c_i= s_{i,\min}+2x_{i,\max}+V(\omega_i\mu+ e_{\max})+\epsilon_1$ where $\epsilon_1\ge 0$.
 For the design of $V_{\max}$, to make the proof in Case $1$ work, it is sufficient to let $V_{\max}= \min_{1\le i \le N}\Big\{\frac{s_{i,\max}-s_{i,\min}-3x_{i,\max}-\epsilon_1-\epsilon_2}{2(\omega_i\mu+ e_{\max})}\Big\}$ where $\epsilon_2>0$. Based on the  proof in Case $2$,
 $\epsilon_1$ and $\epsilon_2$ are further determined as $0$ and $x_{i,\max}$, respectively, to make $V_{\max}$ as large as possible.

\section{Proof of Theorem \ref{the-iid}}\label{app-theiid}
We first give the following fact, which is a direct consequence of the results in \cite{bkneely}.
\begin{lemma}
There exists a stationary randomized regulation allocation solution $(\mb{x}_{d,t}^{s}, \mb{x}_{u,t}^{s})$ that only depends on the system state $A_t$, and there are
\begin{eqnarray}
\label{xs1}
&&\E[x_{i,t}^s]  = z_{i}^s, \forall i, \textrm{ for some } z_{i}^s\in[0,x_{i,\max}],
\\
\label{xs2}
&&\E[e_t^s]-\sum_{i=1}^N\omega_iU(z_i^s)\le -f_2(\hat{\mb{x}}_{d,t},\hat{\mb{x}}_{u,t},\hat{\mb{z}}_t), \\
\label{xs3}
&&\E[\1_{d,t}C_i(x_{id,t}^s)+\1_{u,t}C_i(x_{iu,t}^s)]\le c_{i,\textrm{up}},\forall i,\textrm{ and }\\
\label{xs4}
&& \E[b_{i,t}^s] = 0, \forall i,
\end{eqnarray}
where  the expectations are taken over the randomness of the system and the randomness of $(\mb{x}_{d,t}^{s}, \mb{x}_{u,t}^{s})$, and $(\hat{\mb{x}}_{d,t},\hat{\mb{x}}_{u,t},\hat{\mb{z}}_t)$ is an optimal solution for \textbf{P3}.
\end{lemma}

1)
For brevity,  define $W_t \define \left(\sum_{i=1}^N\omega_iU(z_{i,t})\right) - e_t$.
 Since  WMRA   minimizes the upper bound in (\ref{dpp}), plug $(\mb{x}_{d,t}^s,\mb{x}_{u,t}^s)$ on the right hand side of (\ref{dpp}) together with $z_{i,t} = z_i^s, \forall t$, we have
\begin{eqnarray}\label{th1}
\Delta({\mb{\Theta}}_t )-V\E\left[\tilde{W}_t |{\mb{\Theta}}_t\right]\le B-Vf_2(\hat{\mb{x}}_{d,t},\hat{\mb{x}}_{u,t},\hat{\mb{z}}_t),
\end{eqnarray}
where (\ref{xs1}), (\ref{xs2}),  (\ref{xs3}), and \eqref{xs4} are used.
Since $\tilde{W}_t\le \sum_{i=1}^N\omega_iU(x_{i,\max})$, from (\ref{th1}),
\begin{align*}
\Delta({\mb{\Theta}}_t )\le D\define B+V\left(\sum_{i=1}^N\omega_iU(x_{i,\max})-f_2(\hat{\mb{x}}_{d,t},\hat{\mb{x}}_{u,t},\hat{\mb{z}}_t)\right).
\end{align*}
Using Theorem 4.1
in \cite{bkneely}, $\E[|H_{i,t}|]$ and $\E[|J_{i,t}|]$ are  upper bounded by $\sqrt{2tD+2\E[L(\mb{\Theta}_0)]}, \forall t$. Hence, the virtual queues
$H_{i,t}$ and $J_{i,t}$  are mean rate stable and the following limit constraints hold.
\begin{eqnarray} \label{pztil}
 &&\lim_{T\to\infty} \frac{1}{T}\sum_{t=0}^{T-1} \E[\tilde{z}_{i,t}] = \lim_{T\to\infty} \frac{1}{T}\sum_{t=0}^{T-1} \E[\tilde{x}_{i,t}],\forall i,\\
 \nonumber
 &&\lim_{T\to\infty} \frac{1}{T}\sum_{t=0}^{T-1} \E\left[\1_{d,t}C_i(\tilde{x}_{id,t}) + \1_{u,t}C_i(\tilde{x}_{iu,t}) \right] \le c_{i,\textrm{up}},
\forall i.
\end{eqnarray}
Since $s_{i,t}$ is bounded under WMRA by Lemma \ref{lem-sbd}, using Lemma \ref{lem-cst3}, we have $\lta\E[\tilde{b}_{i,t}] = 0,\forall i.$
In addition, note that $(\tilde{\mb{x}}_{d,t}, \tilde{\mb{x}}_{u,t})$ is derived under the constraints of
the optimization problems (\textbf{a}), (\textbf{b1}), and (\textbf{b2}). Therefore, we have that
$(\tilde{\mb{x}}_{d,t}, \tilde{\mb{x}}_{u,t})$ is feasible
 for \textbf{P3}, \textbf{P2}, and \textbf{P1}.

2) Taking  expectations of both sides of (\ref{th1}) and summing over $t \in \{0,1, \cdots, T-1\}$ for some $T>1$, we have
\begin{eqnarray}\nonumber
\frac{1}{T}\sum_{t=0}^{T-1}\E[\tilde{W}_t]&\ge& \frac{\E\left[L({\mb{\Theta}}_T )-L({\mb{\Theta}}_0)\right]}{VT}+f_2(\hat{\mb{x}}_{d,t},\hat{\mb{x}}_{u,t},\hat{\mb{z}}_t)-B/V\\
\label{th3}
&\ge& f_2(\hat{\mb{x}}_{d,t},\hat{\mb{x}}_{u,t},\hat{\mb{z}}_t)-B/V-\E[L({\mb{\Theta}}_0)]/VT ,
\end{eqnarray}
where
 (\ref{th3}) holds since  $L({\mb{\Theta}}_T)$ is non-negative.
Also,
\begin{align}
\nonumber
\frac{1}{T}\sum_{t=0}^{T-1}\E[\tilde{W}_t]&=\frac{1}{T}\sum_{t=0}^{T-1}\E\left[\left(\sum_{i=1}^N\omega_iU(\tilde{z}_{i,t})\right) - \tilde{e}_t \right]\\
\label{w1}
&\hspace{-1cm}\le\sum_{i=1}^N\omega_iU\left(\frac{1}{T}\sum_{t=0}^{T-1}\E[\tilde{z}_{i,t}]\right)-\frac{1}{T}\sum_{t=0}^{T-1}\E[\tilde{e}_t], 
\end{align}
where the inequality in (\ref{w1}) is derived
using Jensen's inequality for concave functions. Combining (\ref{th3}) and (\ref{w1}) and taking limits on both sides, there is
\begin{eqnarray}
\nonumber
&&\sum_{i=1}^N\omega_iU\left(\lim_{T\to \infty}\frac{1}{T}\sum_{t=0}^{T-1}\E[\tilde{z}_{i,t}]\right)-\lim_{T\to \infty}\frac{1}{T}\sum_{t=0}^{T-1}\E[\tilde{e}_t]\\
\label{w11}
&\ge& f_2(\hat{\mb{x}}_{d,t},\hat{\mb{x}}_{u,t},\hat{\mb{z}}_t)-B/V\\
\label{w22}
&\ge& f_2(\mb{x}^*_{d,t},\mb{x}^*_{u,t},\mb{z}^*_t)-B/V \\
\label{w2}
&=& f_1(\mb{x}^{\textrm{opt}}_{d,t},\mb{x}^{\textrm{opt}}_{u,t})-B/V,
\end{eqnarray}
where $(\mb{x}^*_{d,t},\mb{x}^*_{u,t},\mb{z}^*_t)$ and   $(\mb{x}^{\textrm{opt}}_{d,t},\mb{x}^{\textrm{opt}}_{u,t})$ are defined in Section \ref{subsec-protran},  \eqref{w11} holds since $\E[L({\mb{\Theta}}_0)]$ is bounded,  \eqref{w22} holds since the feasible set of the optimization variables is enlarged from \textbf{P2} to \textbf{P3}, and \eqref{w2} is true due to Lemma \ref{lem-p1p2}.

Rewrite the objective function of \textbf{P1} under WMRA, \ie, $f_1(\tilde{\mb{x}}_{d,t},\tilde{\mb{x}}_{u,t})$, as
\begin{eqnarray*}
\nonumber
&&\sum_{i=1}^N\omega_iU\left(\lim_{T\to \infty}\frac{1}{T}\sum_{t=0}^{T-1}\E[\tilde{z}_{i,t}]\right)-\lim_{T\to \infty}\frac{1}{T}\sum_{t=0}^{T-1}\E[\tilde{e}_t]\\
&+&\sum_{i=1}^N\omega_iU\left(\lim_{T\to \infty}\frac{1}{T}\sum_{t=0}^{T-1}\E[\tilde{x}_{i,t}]\right)\\
&-&\sum_{i=1}^N\omega_iU\left(\lim_{T\to \infty}\frac{1}{T}\sum_{t=0}^{T-1}\E[\tilde{z}_{i,t}]\right).
\end{eqnarray*}
Due to (\ref{pztil}), the last two terms cancel each other. Hence, by   (\ref{w2}), we have $f_1(\tilde{\mb{x}}_{d,t},\tilde{\mb{x}}_{u,t})\ge f_1(\mb{x}^{\textrm{opt}}_{d,t},\mb{x}^{\textrm{opt}}_{u,t})-B/V $, which completes the proof.

\begin{IEEEbiography}[{\includegraphics[width=1in,height=1.25in,clip,keepaspectratio]{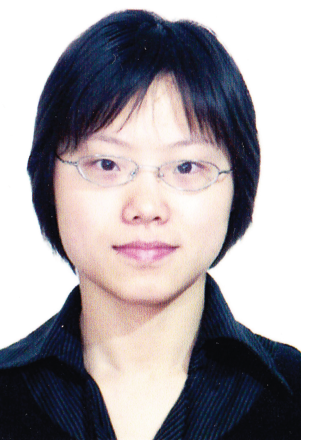}}]
{Sun Sun} (S'11)
received the B.S. degree in Electrical Engineering and Automation  from  Tongji University, Shanghai, China, in 2005. From 2006 to 2008, she was a software engineer in the Department of GSM Base Transceiver Station of Huawei Technologies Co. Ltd.. She received the M.Sc. degree in Electrical and Computer Engineering from University of Alberta, Edmonton, Canada, in 2011.
Now, she is pursuing her Ph.D. degree in the Department of Electrical and Computer Engineering of University of Toronto, Toronto, Canada.
Her current research interest lies in the areas of  stochastic optimization and distributed control, with the application of
energy management in smart grid. 
\end{IEEEbiography}

\begin{IEEEbiography}[{\includegraphics[width=1in,height=1.25in,clip,keepaspectratio]{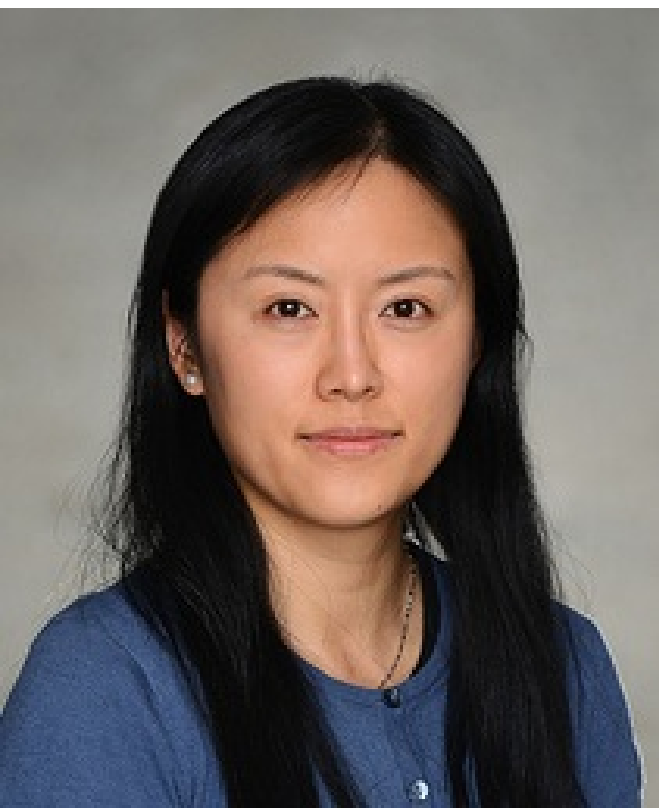}}]
{Min Dong} (S'00-M'05-SM'09) received the B.Eng. degree from Tsinghua University, Beijing, China, in 1998, and the Ph.D. degree in electrical and computer engineering with minor in applied mathematics from Cornell University, Ithaca, NY, in 2004. From 2004 to 2008, she was with Corporate Research and Development, Qualcomm Inc., San Diego, CA. In 2008, she joined the Department of Electrical, Computer and Software Engineering at University of Ontario Institute of Technology, Ontario, Canada, where she is currently an Associate Professor. She also holds a status-only Associate Professor appointment with the Department of Electrical and Computer Engineering, University of Toronto since 2009. Her research interests are in the areas of statistical signal processing for communication networks, cooperative communications and networking techniques, and stochastic network optimization in dynamic networks and systems.

Dr. Dong received the Early Researcher Award from Ontario Ministry of Research and Innovation in 2012, the Best Paper Award at IEEE ICCC in 2012, and the 2004 IEEE Signal Processing Society Best Paper Award. She was an Associate Editor for the IEEE SIGNAL PROCESSING LETTERS during 2009-2013, and currently serves as an Associate Editor for the IEEE TRANSACTIONS ON SIGNAL PROCESSING.  She has been an elected member of IEEE Signal Processing Society Signal Processing for Communications and Networking (SP-COM) Technical Committee since 2013. 
\end{IEEEbiography}

\begin{IEEEbiography}[{\includegraphics[width=1in,height=1.25in,clip,keepaspectratio]{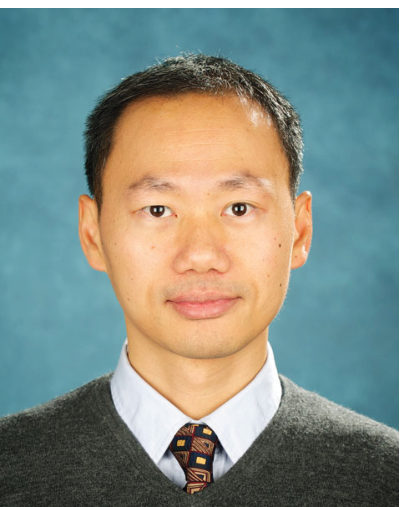}}]
{Ben Liang}(S'94-M'01-SM'06)
received honors-simultaneous B.Sc. (valedictorian) and M.Sc. degrees in Electrical Engineering from Polytechnic University in Brooklyn, New York, in 1997 and the Ph.D. degree in Electrical Engineering with Computer Science minor from Cornell University in Ithaca, New York, in 2001. In the 2001 - 2002 academic year, he was a visiting lecturer and post-doctoral research associate at Cornell University. He joined the Department of Electrical and Computer Engineering at the University of Toronto in 2002, where he is now a Professor. His current research interests are in mobile communications and networked systems. He has served as an editor for the IEEE Transactions on Wireless Communications and an associate editor for the Wiley Security and Communication Networks journal, in addition to regularly serving on the organizational or technical committee of a number of conferences. He is a senior member of IEEE and a member of ACM and Tau Beta Pi. 
\end{IEEEbiography}

\end{document}